\documentclass[twocolumn]{aastex631}
\usepackage{amsmath}
\usepackage{graphicx}
\usepackage{xcolor}
\usepackage{lineno}

\newcommand{\mgfe}[0]{[{\rm Mg/Fe}]} 
\newcommand{\Acc}{A_{\rm cc}}
\newcommand{\AIa}{A_{\rm Ia}} 
\newcommand{\Aagb}{A_{\rm AGB}}

\newcommand{\xmg}{[{\rm X}/{\rm Mg}]} 
 
\newcommand{\xh}{[{\rm X}/{\rm H}]} 
\newcommand{\mgh}{[{\rm Mg}/{\rm H}]}
\newcommand{\feh}[0]{[{\rm Fe/H}]} 
\newcommand{\qIax}{q_{\rm Ia}^{\rm X}}
\newcommand{\qccx}{q_{\rm cc}^{\rm X}}
\newcommand{\qagbx}{q_{\rm AGB}^{\rm X}}
\newcommand{\qIa}{q_{\rm Ia}}
\newcommand{\qcc}{q_{\rm cc}}
\newcommand{\qagb}{q_{\rm AGB}}

\newcommand{\fcc}{f_{\rm cc}}
\newcommand{\logg}{\log(g)}
\newcommand{\teff}{T_{\rm eff}}

\newcommand{\kel}{\rm \; K}

\defcitealias{griffith19}{GJW}
\defcitealias{weinberg2021}{W21}

\submitjournal{ApJ}

\shorttitle{Residual Abunds in GALAH DR3}
\shortauthors{Griffith et al.}
\graphicspath{{./}{}}
\begin{document}

\title{Residual Abundances in GALAH DR3: Implications for Nucleosynthesis and Identification of Unique Stellar Populations}

\correspondingauthor{Emily Griffith}
\email{griffith.802@osu.edu}

\author[0000-0001-9345-9977]{Emily J. Griffith}
\affiliation{The Department of Astronomy and Center of Cosmology and AstroParticle Physics, The Ohio State University, Columbus, OH 43210, USA}

\author[0000-0001-7775-7261]{David H. Weinberg}
\affiliation{The Department of Astronomy and Center of Cosmology and AstroParticle Physics, The Ohio State University, Columbus, OH 43210, USA}
\affiliation{The Institute for Advanced Study, Princeton, NJ, 08540, USA}

\author[0000-0002-4031-8553]{Sven Buder}
\affiliation{Research School of Astronomy \& Astrophysics, Australian National University, ACT 2611, Australia}
\affiliation{Center of Excellence for Astrophysics in Three Dimensions (ASTRO-3D), Australia}

\author[0000-0001-7258-1834]{Jennifer A. Johnson}
\affiliation{The Department of Astronomy and Center of Cosmology and AstroParticle Physics, The Ohio State University, Columbus, OH 43210, USA}

\author[0000-0002-6534-8783]{James W. Johnson}
\affiliation{The Department of Astronomy and Center of Cosmology and AstroParticle Physics, The Ohio State University, Columbus, OH 43210, USA}

\author[0000-0002-0743-9994]{Fiorenzo Vincenzo}
\affiliation{The Department of Astronomy and Center of Cosmology and AstroParticle Physics, The Ohio State University, Columbus, OH 43210, USA}

\begin{abstract}

We investigate the [X/Mg] abundances of 16 elements for 82,910 Galactic disk stars from GALAH+ DR3. We fit the median trends of low-Ia and high-Ia populations with a two-process model, which describes stellar abundances in terms of a prompt core-collapse and delayed Type-Ia supernova component. For each sample star, we fit the amplitudes of these two components and compute the residual $\Delta$[X/H] abundances from this two-parameter fit. We find RMS residuals $\lesssim 0.07$ dex for well-measured elements and correlated residuals among some elements (such as Ba, Y, and Zn) that indicate common enrichment sources. From a detailed investigation of stars with large residuals, we infer that roughly $40\%$ of the large deviations are physical and $60\%$ are caused by problematic data such as unflagged binarity, poor wavelength solutions, and poor telluric subtraction. As one example of a population with distinctive abundance patterns, we identify 15 stars that have 0.3-0.6 dex enhancements of Na but normal abundances of other elements from O to Ni and positive average residuals of Cu, Zn, Y, and Ba. We measure the median elemental residuals of 14 open clusters, finding systematic $\sim0.1-0.4$ dex enhancements of O, Ca, K, Y, and Ba and $\sim0.2$ dex depletion of Cu in young clusters. Finally, we present a restricted three-process model where we add an asymptotic giant branch star (AGB) component to better fit Ba and Y. With the addition of the third process, we identify a population of stars, preferentially young, that have much higher AGB enrichment than expected from their SNIa enrichment.

\end{abstract}


\section{Introduction} \label{sec:intro}

The violent ends of stellar lives bring violent delights. Core-collapse supernovae (CCSN) and Type-Ia supernovae (SNIa) produce the majority of elements from O to Ni in our universe, with each element originating from a unique mix of nucleosynthetic processes. Mg and other $\alpha$-elements, for example, are dominated by CCSN production while Fe-peak elements are produced in both CCSN and SNIa \citep[e.g.,][]{andrews2017, rybizki2017}. These nucleosynthetic processes enrich the interstellar medium with metals that are recycled into the next generation of stars. Since each star bears a chemical fingerprint of the interstellar medium at the time of its birth, we can observe stellar abundances today to learn about the enrichment events of the past. Spectroscopic surveys such as RAVE, SEGUE, LAMOST, \textit{Gaia}-ESO, APOGEE, GALAH, and H3 \citep{steinmetz2006, yanny2009, luo2015, gilmore2012, desilva2015, majewski2017, conroy2019} have reported the abundances of millions of stars in our Galaxy, spanning the disk, halo, and bulge. The GALAH\footnote{GALAH = GALactic Archaeology with HERMES} and APOGEE\footnote{APOGEE = Apache Point Observatory Galactic Evolution Experiment} surveys, in particular, have high spectral resolutions that allow for the determination of over 15 elemental abundances per star, spanning elements produced by multiple enrichment channels. In this paper we focus on abundances from GALAH Data Release 3 \citep[DR3;][]{buder2021} and analyze the population trends as well as the individual stellar measurements to understand our Galactic enrichment history on large and small scales. 

As in our prior works \citep{weinberg2019,griffith2019}, we leverage the bimodal distribution of [$\alpha$/Fe]\footnote{[X/Y] = $\log_{10}({\rm X/Y}) - \log_{10}({\rm X/Y})_{\odot}$} in the solar neighborhood \citep[e.g.,][]{fuhrmann1998, bensby2003, adibekyan2012, vincenzo2021a} to separate stars with the high and low SNIa enrichment. The low-Ia (high-$\alpha$) thick-disk and high-Ia (low-$\alpha$) thin disk arise, in part, from the dominant production of $\alpha$-elements in prompt CCSN, the delayed timescale of SNIa enrichment \citep{maoz2012}, and the significant SNIa enrichment to Fe. While many have studied the two stellar populations in [X/Fe] vs. [Fe/H] space, the use of Mg as a reference element, as advocated by \citet{weinberg2019}, provides a more straightforward interpretation of the abundance trends because Mg has a single enrichment source. In [X/Mg] vs. [Mg/H] abundance space, \citet{weinberg2019} and \citet{griffith2021a} show that while the density of the high-Ia and low-Ia populations varies with Galactic location \citep{nidever2014,hayden2015}, the median  abundance trends, and therefore the implied nucleosynthetic yields, are consistent throughout the disk and bulge.

\citet[][hereafter W21]{weinberg2019,weinberg2021} describe these Galactic abundance trends with the two-process model, which assumes that the abundances of all stars can be described by the sum of a CCSN and SNIa process. From the separation in the population's median high-Ia and low-Ia trends, the two-process model infers the fractional contributions from CCSN and SNIa. We have previously fit the two-process model to large multi-element abundance samples from APOGEE \citep{weinberg2019} and GALAH \citep{griffith2019}, empirically determining the origin of the elements observed by both surveys. In addition to fitting the median trends, the two-process model can predict each star's full set of abundances from a subset of $\alpha$ and Fe-peak elements. With APOGEE data \citetalias{weinberg2021} show that 15 elemental abundances can be accurately predicted from a star's Mg and Fe abundances alone, and they can be more accurately predicted from a fit to six $\alpha$ and Fe-peak elements. 

However, the fit is imperfect, both because of observational errors and because the assumptions of the two-process model are only approximate. The two-process fits allow a star's $N$ abundance measurements to be recast into two parameters that capture the main axes of variation and $N-2$ residuals that traces subtler or rarer deviations from overall trends. \citetalias{weinberg2021} use these residual abundances to characterize enrichment patterns in the APOGEE disk sample. Here we apply a similar approach to GALAH DR3, taking advantage of GALAH's denser sampling of the solar neighborhood and its access to elements that APOGEE does not measure (notably Sc, Zn, Y, and Ba, and more reliable measurements of Ti and Na). 

While most stars are well fit by the two-process model, the residual differences between the observed and predicted abundances hold a wealth of information about the global and local enrichment processes \citep{ting2021}. Residual abundance of individual stars identify interesting enhancements or depletions, contributions from non-CCSN and SNIa processes, and failures of the abundance pipeline. Correlations in abundances residuals of a stellar population hold information on the nucleosyntheitc processes that enrich our Galaxy and their stochasticity. Guided by the conclusions from \citet{ting2021}, \citetalias{weinberg2021} identify groups of elements with positive residual correlations, and stellar populations (e.g., $\omega$-Cen and the Large Magellanic Cloud) with interesting abundance residuals in APOGEE. 

In this paper, we investigate residual abundances in GALAH, complementing \citetalias{weinberg2021}'s work with APOGEE. We compare our results with those from \citetalias{weinberg2021}, and conduct a deeper exploration of interesting stellar populations and stars with the largest residual abundances. 

\citetalias{weinberg2021} estimate that the intrinsic dispersion of two-process residuals is $\sim 0.02-0.03$ dex for most of the well measured APOGEE elements, rising to $\sim 0.06-0.08$ dex for Na, V, and Ce. \cite{ting2021} and \cite{ness2021} find similar values for the intrinsic dispersion of stellar abundances after conditioning on Mg and Fe, which is similar in practice to fitting the two-process model and computing rms residuals. As emphasized in these papers, the correlations of residuals can provide robust evidence of underlying structure in the element distribution, even when the residuals for any individual star are comparable to the measurement uncertainties. These correlations, the median residuals of selected stellar populations, and the rare but distinctive outlier stars can all provide clues to the sources of this residual structure, which could include additional astrophysical processes (e.g., AGB enrichment), stochastic sampling of the CCSN or SNIa populations, or mixing of populations with different enrichment histories or stellar initial mass functions (IMFs). Errors in abundance measurements can also contribute to correlated residuals or large outliers. Distinguishing physical variations from measurement errors is a challenge in all of these analyses, and in our study here.

In Section~\ref{sec:data} we describe the GALAH survey, its recent data release, and the sample selection for this paper. Section~\ref{sec:abundance} presents the [X/Mg] vs. [Mg/H] abundance trends of the high-Ia and low-Ia populations. Here we compare the median trends from GALAH DR2 with those from DR3, and compare our GALAH trends with those from APOGEE DR17 \citepalias{weinberg2021}. In Section~\ref{sec:twoproc} we summarize the two-process model, fit the model to the GALAH data, and discuss the process vectors and amplitudes. With the two-process model fits, we predict the abundances for our stellar sample in Section~\ref{sec:pred&devs} and present the residual abundance distributions. We identify groups of elements with correlated residuals, evaluate the validity of stellar abundances with the largest residuals, and show example spectra and abundance patterns for stars with interesting abundance trends. In Section~\ref{sec:clusters}, we continue our investigation of interesting residuals, focusing on those of known open cluster members. Section~\ref{sec:agb} extends the two-process model to a restricted three-process model, accounting for AGB enrichment (rather than SNIa enrichment) to Ba and Y. We summarize our findings in Section~\ref{sec:conclusions}.

\section{Data} \label{sec:data}

We employ stellar parameters and abundances from GALAH+ DR3 \citep{buder2021}. The GALAH spectroscopic survey observes in optical wavelengths with the HERMES spectrograph on the Anglo-Australian Telescope \citep{desilva2015, sheinis2015}. GALAH+ DR3 is comprised of three main components—the main GALAH DR3 survey targets, the K2-HERMES survey, and the TESS-HERMES survey. The main GALAH survey observes targets with $12 \leq V \leq 14.3$, $\delta < 10$, and $|b| >10^{\circ}$, and it has significant overlap with \textit{Gaia}. It further extends beyond this magnitude range to include GALAH-bright and GALAH-ultrafaint, which captures targets with magnitudes from 9 to 16 \citep{buder2021}. As their names suggest, the K2-HERMES survey \citep{sharma2019} and TESS-HERMES survey \citep{sharma2018} observe stars in the K2 field and in the TESS Southern Continuous Viewing Zone. GALAH+ DR3 also includes targets from other smaller HERMES surveys, including observations of open clusters \citep{martell2017} and the Galactic bulge. In total, GALAH+ DR3 includes 678,423 spectra for 588,571 stars \citep{buder2021}, which we will hereafter refer to as the GALAH or GALAH DR3 sample.
    
While GALAH DR2 \citep{buder2018} employed The Cannon \citep{ness2015}, a data-driven parameter and abundance pipeline, for their spectral analysis, GALAH DR3 uses Spectroscopy Made Easy \citep[SME,][]{valenti1996, piskunov2017} to determine stellar parameters and abundances. The move away from data-driven analysis improves the stellar labels for stars on the edges of the training data, such as stars with high temperatures and/or low metallicities \citep{buder2021}. With SME, GALAH reports stellar parameters and [X/Fe] abundances for Li, C, O, Na, Mg, Al, Si, K, Ca, Sc, Ti, V, Cr, Mn, Co, Ni, Cu, Zn, Rb, Sr, Y, Zr, Mo, Ru, Ba, La, Ce, Nd, Sm, and Eu, with 1D-NLTE models for H, Li, C, O, Na, Mg, Al, Si, K, Ca, Mn, and Ba \citep{amarsi2020}. For more details on the data reduction pipeline, see \citet{kos2017}, \citet{zwitter2020}, and \citet{buder2021}.

While the addition of so many heavy elements is exciting, \citet{buder2021} caution against the use of Rb, Sr, Zr, Mo, Ru, La, Nd, Sm, and Eu without sufficient inspection of the spectra. The spectral features for these elements, along with Co and V, are frequently blended. At low metallicity, C, Al, and many heavy elements also hit a detection limit threshold. 
All of these elements have absorption features within the GALAH wavelength range in principle. Their line strength varies significantly throughout the parameter space. Heavy elements are, for example, only detectable within relatively few giants at the typical GALAH spectrum quality, whereas the atomic C line only has a detectable line strength for the hottest stars. For our study, we thus have to find a compromise between the number of elements and the number of stars that have detectable elemental abundances. Furthermore, we want to avoid problematic elements close to the detection limit.
In our analysis, we focus on O, Na, Mg, Al, Si, K, Ca, Sc, Ti, Cr, Mn, Fe, Ni, Cu, Zn, Y, and Ba. We discuss C in Appendix~\ref{ap:carbon}, though these results should be interpreted with caution. While Nd and Sm would provide further insight on neutron-capture processes, they are detected in a small faction of stars, making their derived abundance trends susceptible to selection biases.
    
In addition to the main catalog of stellar parameters and abundances, the GALAH collaboration has released value-added-catalogs (VACS) containing \textit{Gaia} eDR3 data \citep{gaia2021, fabricius2021}, Bayesian estimates of ages and distances \citep{sharma2018}, and kinematics. We use the \textit{Gaia} data to search for correlations between abundances and kinematics, but we do not draw any strong conclusions from these results. We employ the age estimates in Sections~\ref{subsec:As},~\ref{subsec:age}, and~\ref{subsec:agb_ap}.

\subsection{Sample Selection} \label{subsec:sample}

We apply a variety of cuts to the full GALAH DR3 catalog to ensure that we have a sample of high-quality data. We exclude all stars with flags on the stellar parameters, [Fe/H], or [Mg/Fe] (requiring \texttt{flag\_sp==0}, \texttt{flag\_fe\_h==0}, \texttt{flag\_mg\_fe==0}). We also require SNR $> 40$ (\texttt{snr\_c2\_iraf > 40}). To ensure that we only have one set of abundances for each star, we remove all repeat observations (\texttt{flag\_repeat==0}).
     
After removing low quality and low SNR data, we define the stellar population that we want to study in $\logg$ and $\teff$. To avoid the effects of correlated abundance errors with $\logg$, as seen in APOGEE \citep{griffith2021a} and GALAH clusters \citep{buder2021}, we focus our study on dwarf and subgiant stars with $\logg > 3.5$ but exclude the remaining cool dwarfs with $\logg > 4.5$. This cut ensures that stars in our sample have reliable $\logg$ values and are nearby, with 99\% of our sample falling within 2 kpc of the sun \citep[distances from GALAH DR3 VAC on BSTEP estimates including ages and distances with latest \textit{Gaia} eDR3 parallaxes][]{sharma2018,buder2021,gaia2021}. We further restrict our sample to stars with temperatures $4200 \kel$ $< \teff < 6700 \kel$. The lower limit removes cool dwarfs that suffer from molecular line blending. The upper value limits systematic trends between abundances and rotational broadening because of intrinsically broader and thus shallower lines. The resulting $\teff$ and $\logg$ range is the most reliable region for GALAH values, as confirmed by \citet{buder2021}, and provides us with a set of reliable abundances measured from well detected lines. We make one final cut in metallicity that restricts our sample to $-0.5 < \mgh < 0.5$. This cut ensures that all elements studied are well populated throughout our metallicity range and  removes low-metallicity stars whose abundances push the detection threshold (e.g., Al). In total, this leave us with a sample of 82,910 stars.

\section{Stellar Abundances} \label{sec:abundance}

We analyze the median abundance trends for our sample of GALAH stars. As in \citet{weinberg2019} and \citet{griffith2019} we divide the sample into high-Ia (low-$\alpha$) and low-Ia (high-$\alpha$) populations based on [Fe/H] and [Mg/Fe] abundances. This division separates the stars with significant SNIa enrichment from those without. Analyzing the abundance trends of both populations informs us on the prompt and delayed nucleosynthesis of each element. Adopting the same division as these earlier papers, we define low-Ia stars as those with:
\begin{equation}
\begin{cases}
\mgfe > 0.12 - 0.13\feh,    & \feh<0 \cr
\mgfe > 0.12,               & \feh>0. \cr
\end{cases}
\label{eq:boundary}
\end{equation}
We classify 89\% of our sample as high-Ia stars. We expect to see this dominance of the high-Ia population since the majority of the stars are in the solar neighborhood \citep{hayden2015}.

\subsection{Trends in GALAH DR2 and DR3}\label{subsec:galah_abund}

\begin{figure*}[!htb]
    \centering
    \includegraphics[width=\textwidth]{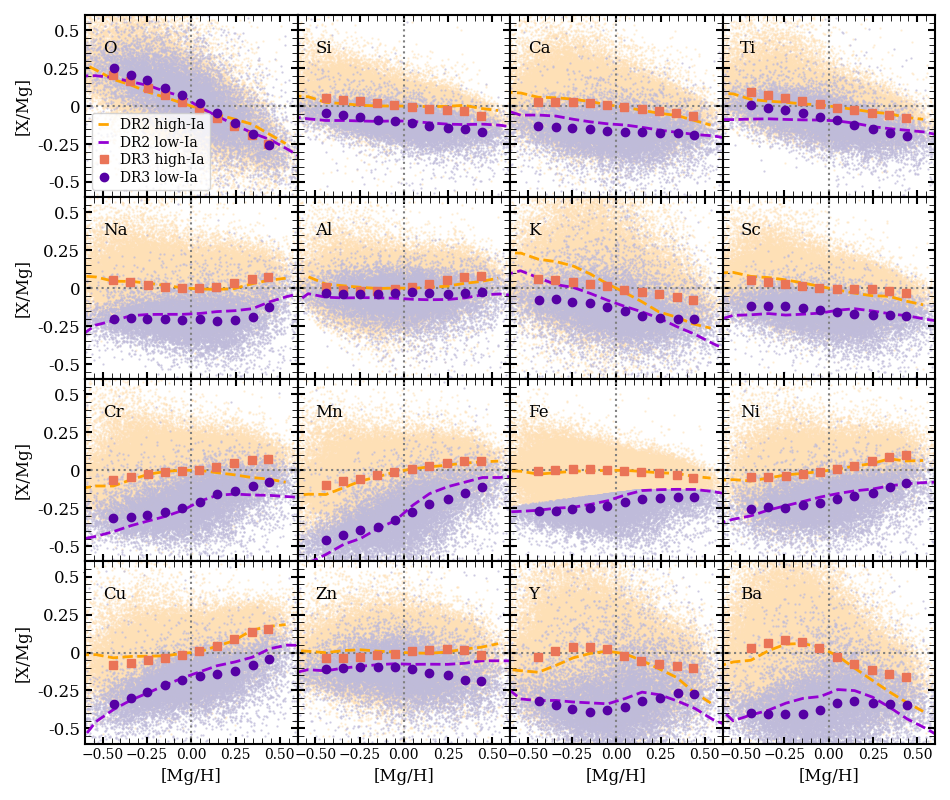}
    \caption{[X/Mg] abundances for GALAH DR3 stars. High-Ia stars are in light orange and low-Ia stars are in light purple. Median high-Ia and low-Ia [X/Mg] abundances trends binned by 0.1 dex are plotted as dark orange squares and dark purple circles, respectively. We include the median high-Ia and low-Ia trends from GALAH DR2 \citep{griffith2019} as orange and magenta dashed lines for comparison.}
    \label{fig:xmgs}
\end{figure*}

In Figure~\ref{fig:xmgs}, we present the elemental abundance distributions of 83,000 stars in [X/Mg] vs. [Mg/H] space for $\alpha$ (O, Si, Ca, Ti), light odd-$Z$ (Na, Al, K, Sc), Fe-peak (Cr, Mn, Fe, Ni), Fe-cliff\footnote{We define “Fe-cliff” as elements on the steeply falling edge of the Fe-peak.} (Cu, Zn), and neutron-capture (Y, Ba) elements. We remove all stars flagged for [Fe/H] or [Mg/Fe] in our full sample selection (Section~\ref{subsec:sample}) and further remove stars with flagged [X/Fe] abundances (\texttt{flag\_x\_fe==0}) in the analysis of each element. We list the number of unflagged stars in our sample for each element in Table~\ref{tab:elems}.
    
\begin{table}[]
\centering
\caption{The number of unflagged stars, the applied zero-point offset in [X/Mg], and the $\fcc$ value inferred along high-Ia sequence at solar metallicity for all elements studied (see Equation~\ref{eq:fcc}).} \label{tab:elems}
\begin{tabular}{cccc}
\tablewidth{0pt}
\hline
\hline
Element & Number & Offset & $\fcc$\\
\hline
O & 80889 & -0.013 & 1.13 \\
Si & 82362 & -0.009 & 0.73 \\
Ca & 81159 & -0.037 & 0.6 \\
Ti & 78512 & -0.023 & 0.78 \\
Na & 82889 & -0.048 & 0.53 \\
Al & 78811 & -0.033 & 0.91 \\
K & 80051 & -0.024 & 0.66 \\
Sc & 82859 & -0.058 & 0.63 \\
Cr & 81324 & 0.051 & 0.50 \\
Mn & 82818 & -0.006 & 0.37 \\
Fe & 82910 & -0.008 & 0.51 \\
Ni & 67826 & 0.030 & 0.53 \\
Cu & 75491 & -0.031 & 0.6 \\
Zn & 79279 & -0.020 & 0.73 \\
Y & 82585 & 0.029 & 0.29 \\
Ba & 82832 & 0.002 & 0.31 \\
\hline
\end{tabular}
\end{table}

We calculate the median [X/Mg] values of the high-Ia and low-Ia populations for each element in bins of 0.1 dex in [Mg/H]. The metallicity cut applied in Section~\ref{subsec:sample} was chosen such that all bins for all elements have $>20$ stars. To ensure that stars on the high-Ia sequence have a solar [X/Mg] abundance at solar [Mg/H], we add a single [X/Mg] offset to all sample stars such that the median high-Ia trend passes through zero, as done in \citet{griffith2019} and \citetalias{weinberg2021}. Our offsets are applied in addition to those in GALAH. The zero points for GALAH DR3 are estimated via abundances of solar (skyflat) spectra and adjusted where needed based on the comparison with stars of the solar circle and solar twins \citep[see][for details]{buder2021}. All our offsets are within 0.06 dex compared to theirs and are reported in Table~\ref{tab:elems}.

We plot the GALAH DR3 median [X/Mg] vs. [Mg/H] trends of the high-Ia and low-Ia populations as solid points in Figure~\ref{fig:xmgs} with the GALAH DR2 medians \citep{griffith2019} for comparison. We show $\alpha$-elements (O, Si, Ca, Ti) in the top row, light odd-$Z$ elements (Na, Al, K, Sc) in the second row, Fe-peak elements (Cr, Mn, Fe, Ni) in the third row, and Fe-cliff (Cu, Zn) and neutron-capture elements (Y, Ba) in the final row. This abundance group separation will continue throughout the paper.
    
We find good agreement between the GALAH DR2 and DR3 high-Ia and low-Ia medians for the $\alpha$-elements, especially O and Si. We see $\sim 0.1$ dex changes in the low metallicity end of the low-Ia trends for Ca and Ti. As in GALAH DR2 and other optical studies, we observe a strong metallicity dependence in the [O/Mg] abundances. O shows no separation between the high-Ia and low-Ia medians, as expected if both Mg and O come purely from CCSN. However, a sloped trend (which is not seen in APOGEE) would require a metallicity dependence of the relative IMF-averaged yields of O and Mg in this regime, surprising as they are expected to arise in similar stars. Si, Ca, and Ti show some separation and thus some contribution from SNIa, in agreement with supernova yield predictions from \citet{andrews2017} and \citet{rybizki2017}, who draw qualitatively similar conclusions about the relative predicted contributions of CCSN, SNIa, and AGB enrichment to different elements. We find that Ca has the largest separation between the high-Ia and low-Ia median trends of the $\alpha$-elements, implying the largest relative SNIa contribution.

Na, K and Sc also show significant separation in their high-Ia and low-Ia medians. While the [Na/Mg] and [Sc/Mg] medians agree between GALAH DR3 and DR2, we see that the [K/Mg] trends have increased sequence separation in DR3 and a significantly flatter slope. These changes are likely due to new NLTE corrections in GALAH DR3 that improve the reliability of the K abundances \citep{buder2021}. K now strongly resembles Sc, another light odd-$Z$ element with similar nucleosynthetic origins \citep{andrews2017}. Unfortunately, K still suffers from interstellar contamination, which would artificially inflate the measured K abundances. The [Al/Mg] medians show little to no separation, differing from the other light odd-$Z$ elements. We see that the separation between the Al medians has decreased slightly from DR2, potentially caused by adjustments to the applied NLTE correction \citep{buder2021}. The close high-Ia and low-Ia medians suggest that Al is dominated by CCSN production.

Unlike the lighter elements, those on the Fe-peak have significant, and often dominant, production in SNIa \citep{andrews2017}. All four Fe-peak elements display large separation between the high-Ia and low-Ia medians. Mn shows the largest separation of all elements. We find similar separation of the high-Ia and low-Ia medians in the DR3 and DR2 abundances for all Fe-peak elements. The [Cr/Mg] trends differ the most with median separation at super-solar metallicities in DR3. We observe small ($\lesssim 0.1$ dex) differences in the low metallicity tails of the low-Ia medians for Cr, Mn, and Ni, with the DR3 trends appearing flatter than those from DR2. 
    
Among the Fe-cliff elements, Cu resembles Mn with steeply sloped median trends and significant separation between the medians. The trends agree well with those from DR2. Relative to Cu, the [Zn/Mg] medians are much flatter and show less separation, though there is a growing gap at high metallicity, diverging from the DR2 trends. The Zn absorption lines are located in heavily blended regions of the blue wavelength region. The numerous absorption features in this wavelength region further complicate normalisation. GALAH DR3 Zn abundances are more trustworthy than those in DR2 thanks to improved normalisation routines and line-by-line measurements, but a significant scatter in the Zn abundances remains.

The Y and Ba medians display different metallicity dependence than the lighter elements. Here, the median high-Ia trends peak near [Mg/H]$\approx -0.25$, and the low-Ia trends incline to a potential peak at super solar metallicities. Because these elements are formed through neutron capture, the expected abundance trends depend upon the availability of seeds and free neutrons. At low metallicity, the abundance of Y and Ba increases as the number of seeds increases. The abundances grow and turn over when the the seed to neutron ratio drops too low to produce Y and Ba \citep{gallino1998}. \citet{griffith2019} show that the high-Ia and low-Ia abundance peaks of both elements align in [Fe/H] space, supporting the theory that Fe-peak elements provide the seeds for these elements \citep{kappeler2011}. We see similar behavior of the [Y/Mg] and [Ba/Mg] trends in DR3 as in DR2, though the peaks in the low-Ia medians are less defined. These differences likely come from the difference in abundance analysis, as the data-driven method implemented in GALAH DR2 may have improperly imposed trends on the neutron-capture element abundances. In both GALAH DR2 and DR3, Y and Ba show large separation, indicative of a strong delayed component that likely originates from AGB stars rather than SNIa \citep{arlandini1999, bisterzo2014}. 

\subsection{Comparison with APOGEE} \label{subsec:apogee}

\begin{figure*}[!htb]
    \centering
    \includegraphics[width=\textwidth]{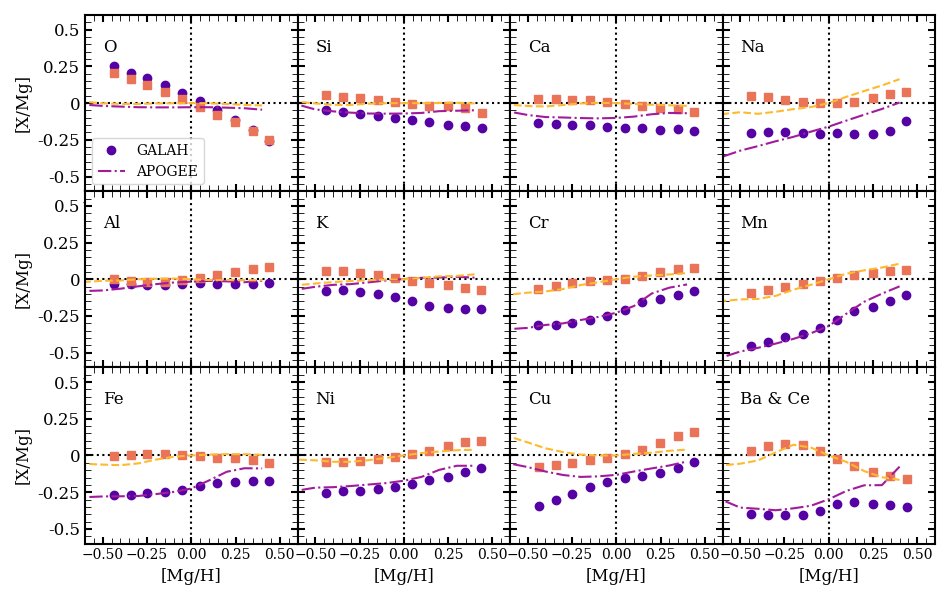}
    \caption{Median [X/Mg] trends the high-Ia (dark orange squares) and low-Ia (dark purple circles) for GALAH DR3 abundances (same as Figure 1). We compare these median trends to those from APOGEE DR17 (light orange and magenta dashed lines, \citetalias{weinberg2021}). We include Cu from APOGEE DR16 and compare GALAH Ba with APOGEE Ce.}
    \label{fig:xmgs_apogee}
\end{figure*}

Comparing the GALAH DR3 median abundance trends to those of APOGEE DR17 \citep[][SDSS Collaboration, in prep]{majewski2017} can highlight which results are robust and identify interesting discrepancies. Though the two surveys observe in different wavelengths, with GALAH in optical and APOGEE in infrared, they have significant overlap in the elements that they measure. In Figure~\ref{fig:xmgs_apogee} we plot the median abundance trends from our GALAH DR3 sample (same as Figure~\ref{fig:xmgs}) with the APOGEE DR17 median abundance trends from \citetalias{weinberg2021}, who select a population of disk ($3 < R< 13$ kpc, $|Z|<2$ kpc) giants ($1 < \logg < 2.5$ and $4000<\teff<4600$ K) with high SNR. Both samples are binned by 0.1 dex in [Mg/H] and zero-point shifted. We include [Cu/Mg] median trends from a similar sample of APOGEE DR16 stars since Cu is excluded in the latest data release. 

The largest difference between GALAH and APOGEE at all metallicities is for O. While the high-Ia and low-Ia [O/Mg] medians show no separation in either survey, the GALAH trends have a steeply decreasing metallicity dependence while the APOGEE trends are flat. This difference is observed between most optical and near-IR O abundances \citep[e.g.,][]{bensby2014} and may arise due to 3D NLTE effects in the optical O triplet \citep[e.g.,][]{kiselman1993, amarsi2020} or systematics in modeling the molecular effects in the IR CO and OH lines \citep[e.g.,][]{collet2007, hayek2011}. 

For Si, Ca, Cr, Mn, Fe, and Ni, there is good agreement below [Mg/H]=0 but disagreement in the super-solar low-Ia trends. For all six elements we see smaller separation between the APOGEE high-Ia and low-Ia medians at high [Mg/H] than those from GALAH. This could be partially explained by a larger SNIa contribution in the APOGEE sample than in GALAH due to sample selection and population cuts. We have inspected a population of dwarf stars observed by both surveys and find that higher metallicity stars classified in as low-Ia in GALAH (Equation~\ref{eq:boundary}) have lower [Mg/Fe] abundances in APOGEE and overlap with the high-Ia population. The difference in GALAH and APOGEE abundances for the same stars suggests that observational uncertainties may be causing the two populations to entangle themselves at super-solar [Mg/H], though there may also be other complications. Further work will be required to understand the origin of the disagreement in the measured abundances of the surveys' overlapping population. While this investigation is outside the scope of our paper, we remain cautions in interpreting the differences in the GALAH and APOGEE low-Ia medians at high [Mg/H].

Two of the three overlapping light odd-$Z$ elements exhibit poor agreement with the APOGEE trends. While both GALAH and APOGEE find significant separation in the [Na/Mg] high-Ia and low-Ia medians, they show different metallicity dependence, with the APOGEE trends rising more steeply than those from GALAH. Na is difficult to observe in APOGEE, but the GALAH Na measurements are robust in dwarfs and subgiants, so we have greater trust in the GALAH metallicity dependence. The similar degree of separation, however, affirms our conclusion in \citet{griffith2019} that Na has a significant delayed contribution, contrary to the theoretical expectations from \citet{andrews2017}. 

\citetalias{weinberg2021} find no separation between the [Al/Mg] and [K/Mg] median trends in APOGEE. This is in good agreement with the Al trends from GALAH, but strong disagreement with K. The [K/Mg] medians differ in both sequence separation and metallicity dependence between the two surveys, with the declining GALAH medians showing more separation than the flat APOGEE trends. We interpret these differences with caution, as K abundances have high uncertainties in APOGEE and may be skewed by interstellar contamination in GALAH.

Finally, we compare the APOGEE and GALAH trends for two heavier elements. While Cu was added to the APOGEE DR16 catalog \citep{jonsson2020}, it suffers from poor detection at low metallicity and was removed in DR17. We include the APOGEE DR16 median Cu trends for the same population cuts as in \citetalias{weinberg2021}. While we see obvious disagreement at low metallicity, where the APOGEE trends may be skewed by blending in weak Cu features, the two surveys show similar separation between their high-Ia and low-Ia medians above [Mg/H] = 0, indicative of a large delayed contribution to Cu. APOGEE DR17 adds Ce, a neutron-capture element near Ba on the periodic table ($Z_{\text{Ce}}=58$ and $Z_{\text{Ba}}=56$) with a similar level of $s$-process contribution \citep[$\sim75-85\%$;][]{arlandini1999, bisterzo2014}. We plot the median [Ba/Mg] median abundances from GALAH DR3 with the median [Ce/Mg] abundances from APOGEE DR17 in the final panel of Figure~\ref{fig:xmgs_apogee}. We see an almost identical peak in the high-Ia median trends at [Mg/H]$\approx -0.2$ and good agreement in the low-Ia medians below [Mg/H] = 0.1. The agreement in metallicity dependence is expected given the two elements' similar atomic numbers and $s$-process enrichment.

\section{Two-Process Model} \label{sec:twoproc}

The two-process model was developed by \citet{weinberg2019} and \citetalias{weinberg2021} to separate and describe the contribution of CCSN and SNIa to elemental abundances. The model assumes that every element is produced by some combination of one delayed source (SNIa) and one prompt source (CCSN), ignoring contributions from other sources such as AGB stars. While the original model was restricted to two processes with power law metallicity dependences, \citetalias{weinberg2021} introduce a revised two-process model that can reproduce any metallicity dependence and can be extended to include additional components. 

Here we will employ the two-process model with only a CCSN and SNIa process, though we will consider an AGB process later in Section~\ref{sec:agb}. The model describes every star through a combination of the two nucleosynthetic sources, or components. Each component consist of two parts: a CCSN/SNIa process vector ($\qccx(z)$ or $\qIax(z)$, where $z\equiv10^{\rm [Mg/H]}$), specific to each element at each $z$ but constant for all stars, and a CCSN/SNIa amplitude ($\Acc$ or $\AIa$), specific to each star but constant for all elements. We define $\Acc=\AIa=1$ for a star with solar abundances ([X/H]=0 for all X).

Together, these components describe [X/H] and [X/Mg] through vector addition as
    \begin{equation}\label{eq:xh}
        \xh = \log_{10}[\Acc\qccx(z) + \AIa\qIax(z)]
    \end{equation}
and
    \begin{equation}\label{eq:xmg}
        \xmg = \log_{10}[\qccx(z)+\qIax(z)\AIa/
        \Acc]
    \end{equation}
where $z \equiv 10^{\mgh}$.
We can further describe the fractional CCSN contribution to $X$ with these parameters, such that
    \begin{equation}\label{eq:fcc}
        \fcc^{X} = \frac{\Acc\qccx(z)}{\Acc\qccx(z) + \AIa\qIax(z)}.
    \end{equation}
    
To infer the values of $\qccx(z)$ and $\qIax(z)$ from observed median sequences, we make the following key assumptions:
    \begin{enumerate}
        \item Mg is a pure CCSN element ($q_{\rm Ia}^{\rm Mg} = 0$).
        \item The Mg and Fe processes are independent of metallicity ($q_{\rm cc}^{\rm Mg}(z) = q_{\rm cc}^{\rm Mg}$, $q_{\rm cc}^{\rm Fe}(z) = q_{\rm cc}^{\rm Fe}$, and $q_{\rm Ia}^{\rm Fe}(z) = q_{\rm Ia}^{\rm Fe}$).
        \item The low metallicity [Mg/Fe] abundance of the low-Ia stellar population plateaus at [Mg/Fe] $= 0.3$ (see Figure~\ref{fig:xmgs}).
        \item Stars on the plateau only have Fe enrichment from CCSN ($\AIa = 0$).
    \end{enumerate}
With these assumptions, we can express the process vectors $\qIax$  and $\qccx$ for each element in terms of the high-Ia and low-Ia median [X/Mg] and [Fe/Mg] abundances and the value of the [Mg/Fe] plateau. For a full derivation of the process vectors, refer to Section 2 of \citetalias{weinberg2021}. We describe our process vectors and process amplitudes in Sections~\ref{subsec:qs} and~\ref{subsec:As}, respectively, and show example two-process predicted abundances for simplified cases in Figure~\ref{fig:fccs}. 


\subsection{Process Vectors} \label{subsec:qs}

\begin{figure*}[!htb]
    \centering
    \includegraphics[width=\textwidth]{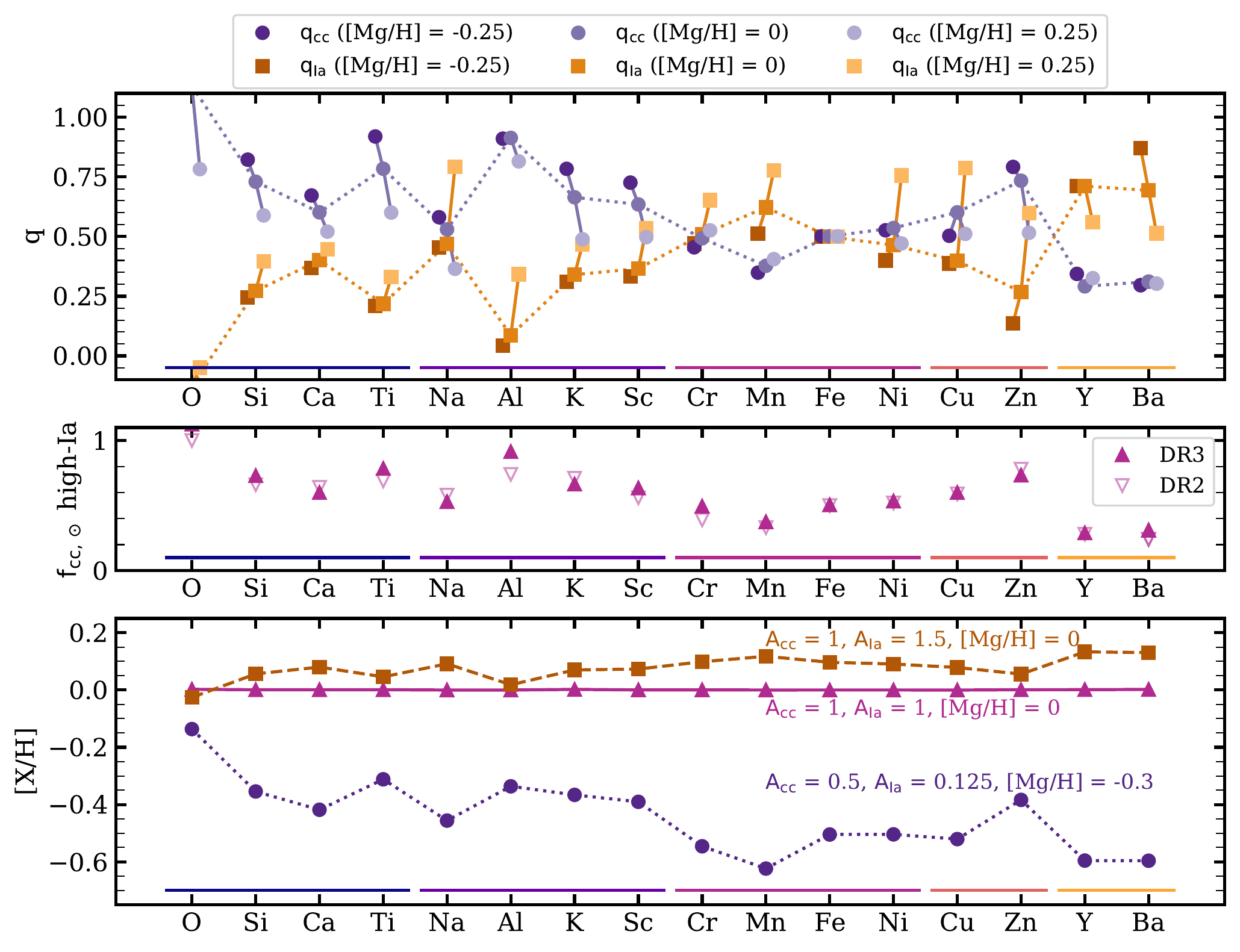}
    \caption{\textit{Top:} Process vectors $\qccx$ (orange squares) and $\qIax$ (purple circles) for all elements at [Mg/H] of -0.25, 0.0, and 0.25. We connect the $\rm{q}$s for each element with solid lines and connect the solar metallicity qs for all elements with dotted lines to guide the eye. \textit{Middle:} The fractional contribution to an element from CCSN derived from the high-Ia median sequence at solar metallicity. $\fcc$ values from this paper (solid triangles) can be compared with those from GALAH DR2 \citep[open triangles,][]{griffith2019}. \textit{Bottom:} Example [X/H] abundances calculated with the GALAH DR3 process vectors. The pink triangles show the abundances of a star at solar metallicity with $\Acc=\AIa=1$. The orange squares represent a high-Ia star at solar [Mg/H] with $\Acc=1$ and $\AIa=1.5$. The purple circles represent a low-Ia star at [Mg/H]=-0.3 with $\Acc=0.5$ and $\AIa=0.125$. The high-Ia and low-Ia abundance trends can be compared to the $\qccx$ and $\qIax$ vectors in the top panel.  Colored bars group elements with common physical characteristics. }
    \label{fig:fccs}
\end{figure*}

We fit the median high-Ia and low-Ia trends with the two-process model as described above, deriving $\qccx$ and $\qIax$ for each [X/Mg] vs. [Mg/H] bin. All $\qccx$ and $\qIax$ values are reported in Tables~\ref{tab:qs1} and~\ref{tab:qs2}, and can be used to reconstruct the high-Ia and low-Ia medians. In the top panel of Figure~\ref{fig:fccs}, we plot $\qccx$ and $\qIax$ for all elements at [Mg/H] = -0.25, 0.0, and 0.25. Each element's $\qccx$ and $\qIax$ values are connected with a solid line to help identify elemental metallicity trends. The solar metallicity $\qccx$ and $\qIax$ for all elements are connected with a dotted line to view the process dependence on element group/atomic number. 

By definition, we find $\qccx+\qIax = 1$ at [Mg/H]$=0$. Elements dominated by CCSN production will have high values of $\qccx$ at all metallicities (e.g., O), while those with significant SNIa enrichment show $\qccx$ and $\qIax$ of more comparable values (e.g., Na). The metallicity dependence of CCSN or SNIa yields is shown by the inclination of the process vectors with increasing [Mg/H] (e.g., Mn). 

In the middle panel, we plot the fraction of each element inferred to come from CCSN ($\fcc$) at solar metallicity from the high-Ia population, also included in Table~\ref{tab:elems}. We stress that the $\fcc$ values are not universal, but are specific to each star or bin of stars; for example, a star on the low-Ia plateau has $\fcc^{\rm Fe}=1$ because it has no SNIa enrichment, even though Fe has a large SNIa contribution in the sun. We plot the values derived in this paper alongside those from GALAH DR2 \citep{griffith2019}. Overall, we find good agreement in elemental $\fcc$ values between the two data releases. Small differences, such as an increased $\fcc$ for Al, follow from the differences in the median abundance trends observed in Figure~\ref{fig:xmgs}. 
In the special case where the high-Ia median crosses below the low-Ia median, as occurs for O, the two-process model produces negative values of $\qIax$ and $\fcc$ values above 1. These values are unphysical and should instead be viewed as $\qIax \approx 0$ and $\fcc \approx 1$.

We find that the $\alpha$-elements, O, Si, Ca, and Ti, have  $\qccx$ and $\fcc$ values above 0.5 at all metallicities. O, which is theoretically expected to be a nearly pure CCSN element, has $\qIax \approx 0$ and $\fcc \approx 1$ at all metallicities. CCSN dominate the production of Si, Ca, and Ti, with Ti having the highest $\fcc$ of the three (0.78) and Ca the lowest (0.60). Similarly, we find that Ca has a weaker CCSN contribution and stronger SNIa  contribution at all metallicities than Si and Ti. All $\alpha$-elements show $\qccx$ decreasing with [Mg/H], in accord with the declining median trends in Figure~\ref{fig:xmgs}, and they show little metallicity dependence in the $\qIax$ vectors. 
 
The process vectors of light odd-$Z$ elements Al, K, and Sc follow a similar metallicity dependence to the $\alpha$-elements, though they all exhibit larger increases in $\qIax$ at high metallicity. Na shows the strongest SNIa process of the light odd-$Z$ elements and, like Al, has a strongly rising SNIa component at high [Mg/H], with $\qIa^{\text{Na}} \sim 0.9$ at [Mg/H]=0.45. We find that Al is almost entirely produced in CCSN at solar metallicity, with the second highest $\fcc$ of the elements studied here ($\fcc = 0.89$). This CCSN fraction agrees with theoretical yields \citep[e.g.,][]{andrews2017} better than the $\fcc = 0.74$ found by \citet{griffith2019}, a change that follows from the observed decrease in the [Al/Mg] median trend separation in GALAH DR3, relative to DR2. 

All Fe-peak elements show a strong SNIa process contribution, especially at high [Mg/H]. By definition, $\qIax = \qccx = 0.5$ for Fe at all metallicities. While the Fe processes are metallicity independent by construction, we see a strong metallicity dependence in the SNIa process for Cr, Mn, and Ni, as the $\qIax$ vectors grow with increasing [Mg/H]. We find no metallicity dependence in the Ni $\qcc$ vectors and a weak metallicity dependence in that of Cr and Mn. Cr and Ni have $\fcc \approx 0.5$ at [Mg/H]$ = 0$. Mn has the largest SNIa contribution of all elements studied here, with $\fcc = 0.38$. 

The Fe-cliff elements Cu and Zn show a strong, positive metallicity dependence in their SNIa process vectors above [Mg/H] = 0, similar to Na and Al. The yield models of \citet{andrews2017} predict that both elements are mainly produced by CCSN, but we infer significant delayed contribution to both that may come from SNIa \citep[e.g.,][]{lach2020} or AGB \citep[e.g.,][]{karakas2016}. At solar metallicity we find $\fcc=0.60$ for Cu and $\fcc=0.73$ for Zn.

As discussed in Section~\ref{subsec:galah_abund}, the neutron-capture elements display a unique metallicity dependence in their high-Ia and low-Ia medians. This translates to their process vectors, which display $\qIax$ that peaks at intermediate [Mg/H], qualitatively resembling the shape of the high-Ia median trends. Both elements have $\qIax \gtrsim 0.5$ at all metallicities, where the $\qIax$ process represents the delayed component, likely AGB stars. Both Y and Ba have almost constant values of $\qccx \approx 0.3$ at all metallicity. We include $\fcc$ for Y and Ba in Figure~\ref{fig:fccs}, but these values should be cautiously interpreted because our separation of prompt and delayed components implicitly assumes that the delayed component tracks SNIa Fe. The prompt (massive star) contribution is expected to be $r$-process, while the delayed (AGB) contribution is expected to bs $s$-process, though we note that the ``$r$'' and ``$s$'' in these two terms refer to the speed of neutron capture relative to $\beta$-decay and not to the rapidity of enrichment relative to star formation. We find $\fcc \approx 0.3$ for Y and Ba, in agreement with results from \citet{arlandini1999} and \citet{bisterzo2014} $s$-process in AGB stars dominates production of both elements.

\subsection{Process Amplitudes} \label{subsec:As}

After calculating $\qccx$ and $\qIax$ from the high-Ia and low-Ia medians for each element, we determine the best-fit process amplitudes for each star in our sample. In \citet{weinberg2019} and \citet{griffith2019} we only employed the Mg and Fe abundances in the $\Acc$ and $\AIa$ calculation. Consequently, if a star's Mg or Fe fluctuated high or low, its amplitudes would be misrepresentative of the other $\alpha$ or Fe-peak elements, a phenomenon referred to as ``measurement aberration'' by \citet{ting2021}. To minimize these effects, we infer $\Acc$ and $\AIa$ from a weighted fit to six trusted elements: Mg, Si, Ca, Ti, Fe, Ni. We iteratively determine the amplitudes for each star such that we minimize the $\chi^2$ value of the fit to these six elements. \citetalias{weinberg2021} find that fitting to six elements (they use Mg, O, Si, Ca, Fe, and Ni) greatly reduces the impact of measurement aberration on the correlation of residual abundances. 

The value of $\Acc$ provides a measurement similar to metallicity (specifically [Mg/H], but with a linear scale), and $\AIa/\Acc$ provides a measure similar to [Fe/Mg]. At solar abundances, $\Acc = \AIa = 1$. With the process vectors for each element and process amplitudes for each star we can calculate $\xh$ according to Equation~\ref{eq:xh}. We plot three example cases of this vector addition in the bottom panel of Figure~\ref{fig:fccs} (as in Figure 3 of \citetalias{weinberg2021}). All take the $\qccx$ and $\qIax$ values derived from the GALAH data. The pink triangles show the case of a star with $\Acc = \AIa = 1$. All [X/H] abundances are solar by construction. The orange squares and purple triangles plot example high-Ia and low-Ia stars, respectively. The high-Ia star has $\Acc =1$ and $\AIa=1.5$, resulting in super solar abundances of all elements, more so for elements with large $\qIax$. Conversely, the abundance pattern of the low metallicity, low-Ia star ($\Acc = 0.5$ and $\AIa = 0.125$) resembles a scaled version of the $\qccx$ vector, with small augmentation of elements with high $\qIax$.

\begin{figure*}[!htb]
    \centering
    \includegraphics[width=\textwidth]{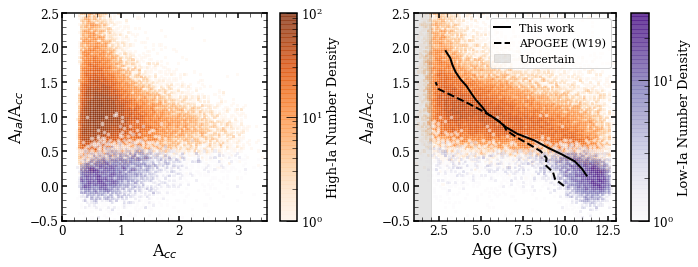}
    \caption{\textit{Left:} The CCSN and SNIa process amplitudes plotted as $\AIa/\Acc$ vs. $\Acc$. High-Ia stars are in orange and low-Ia stars are in purple, with the shade indicating the number density of stars in each cell. This figure resembles the [Fe/Mg] vs. [Mg/H] plot. \textit{Right:} $\AIa/\Acc$ vs. age, with stellar ages estimated from BSTEP. We plot the median age in bins of $\AIa/\Acc$ for our sample (solid line) and from APOGEE (dashed line, \citetalias{weinberg2021}). We shade the region from 0 to 2 Gyrs grey to indicate uncertain ages, as those below and near 1 Gyr may be overpredicted or missing because of our sample selection (Section~\ref{subsec:sample})}
    \label{fig:As}
\end{figure*}

We assign each star a best-fit $\Acc$ and $\AIa$ to predict the full suite of abundances, resembling the examples in Figure~\ref{fig:fccs}. We plot the distribution of the $\AIa/\Acc$ vs. $\Acc$ values for our stellar sample in the left panel of Figure~\ref{fig:As}. This plot can be read like the Tinsley-Wallerstein diagram ([Mg/Fe] vs. [Fe/H],  \citealp{wallerstein1962, tinsley1979, tinsley1980}) such that the high-Ia and low-Ia populations separate. The minimum value of $\AIa/\Acc \approx 0.5$ for the low-Ia population in this diagram largely follows from our definition of this population (Equation~\ref{eq:boundary}). The median ratio rises slowly with $\Acc$, tracing the rise of [Fe/Mg] above the plateau at -0.3. 

Relative to APOGEE (see \citetalias{weinberg2021}, Figure 8), the GALAH stars show a tail of values up to $\AIa/\Acc \approx 2.5$, while the APOGEE ratios cut off at 1.5. In principle this difference could arise from a difference in samples, but this seems unlikely because the APOGEE distribution is consistent throughout the disk.
Instead, it probably arises from differences in the abundance measurements, likely from scatter in the GALAH abundances. We find that 1698 stars have $\AIa/\Acc > 2.25$. Of these, 93\% have low [Mg/H] ($<-0.1$) relative to other $\alpha$-elements, suggesting a measurement error in this abundance. We inspected the spectra of stars with high $\AIa/\Acc$ and low [Mg/H], and noted clear signatures of rotational broadening. Roughly 75\% of stars with $\AIa/\Acc > 2.25$ and [Mg/H]$<-0.1$ have $v_{\rm broad}> 20$ km/s (as fit by SME). Because GALAH reports low [Mg/H] and high [Fe/Mg] ($>0.1$) abundances for these stars, the two-process model fits them with a low $\Acc$ to reproduce the low Mg and a high $\AIa$ to compensate for higher Ca and Si abundances, since both have an SNIa component. This results in an overprediction of Fe and a high $\AIa/\Acc$. A total of 5973 stars in our sample ($\sim7\%$) have broadening velocities greater than 20 km/s. We repeated the prior components of our analysis excluding these fast rotators and found no significant changes to the median abundance trends or process vectors. Their exclusion does reduce the density of stars in the high $\AIa/\Acc$ tail observed in Figure~\ref{fig:A_dist}, but it does not remove all stars with a high amplitude ratio.

As the relative CCSN and SNIa contributions change with time, we also plot $\AIa/\Acc$ vs. age in the right hand panel of Figure~\ref{fig:As}. We use ages derived by the Bayesian Stellar Parameter Estimation code \citep[BSTEP, ][]{sharma2018}, excluding stars whose ages have a fractional error $>25\%$. These age estimates are derived from PARSEC release v1.2S + COLIBRI stellar isochrones \citep{marigo2017} and are provided in the GALAH value-added catalog \texttt{GALAH\_DR3\_VAC\_ages}. We note that our temperature cut of $\teff < 6700$ excludes many young stars from this diagram, and that stellar ages below or near 1 Gyr may be overestimated \citep{sharma2020}.

We find that the low-Ia population is dominated by old stars with ages between 10 and 12 Gyrs. These stars all have low values of $\AIa/\Acc$, and they show little evolution in the amplitude ratio with age. We observe a small tail of low-Ia stars at younger ages, with $15\%$ having age $<8$ Gyrs. The high-Ia population spans the full age range probed by BSTEP, but it has the highest density of stars at ages of 2 to 6 Gyrs. Stars with higher $\AIa/\Acc$ values tend to be younger. The solid curve shows the median age of stars in bins of $\AIa/\Acc$. The dashed curve shows the corresponding trend from APOGEE \citepalias{weinberg2021}, with red giant ages inferred from spectra using a Bayesian neural network \citep{leung2019}. Both analyses find a trend of age with $\AIa/\Acc$ within the high-Ia and low-Ia populations, as well as the difference in typical age between them. The APOGEE sample has few ages beyond 10 Gyrs, most likely because the C/N ratios that are the principal diagnostic saturate at large ages \citep{mackereth2017}. The GALAH trend based on isochrone ages is likely more accurate. At $\AIa/\Acc\approx1.3$ the APOGEE ages are younger, and here it is less obvious which trend is more reliable. The difference in median trends is likely tied to $\AIa/\Acc$ differences and connected to the tail of higher $\AIa/\Acc$ values in GALAH, which may itself be driven by rotation affecting abundances. The majority of stars with $v_{\rm broad}>20$ km/s are young (age $<$ 4 Grs) and have thus had less time to lose angular momentum (see further discussion in Section~\ref{subsec:age}). We compared the median APOGEE and GALAH $\AIa/\Acc$ vs. age trends after excluding the stars with high rotational broadening, again noting the decreased density of stars with high $\AIa/\Acc$. This exclusion did not improve the agreement between the median APOGEE and GALAH ages for $\AIa/\Acc > 1.3$, indicating that GALAH has systematically older stars than APOGEE in this amplitude regime. It is unclear if this is due to differences in the age or amplitude scales. We also have checked for correlations between the process amplitudes and other stellar parameters, such as eccentricity, Galactic location, and kinematics information, but we find no clear trends within the GALAH sample. 

\section{Two-Process Fits and Residual Abundances} \label{sec:pred&devs}

\begin{figure*}[!htb]
    \centering
    \includegraphics[width=\textwidth]{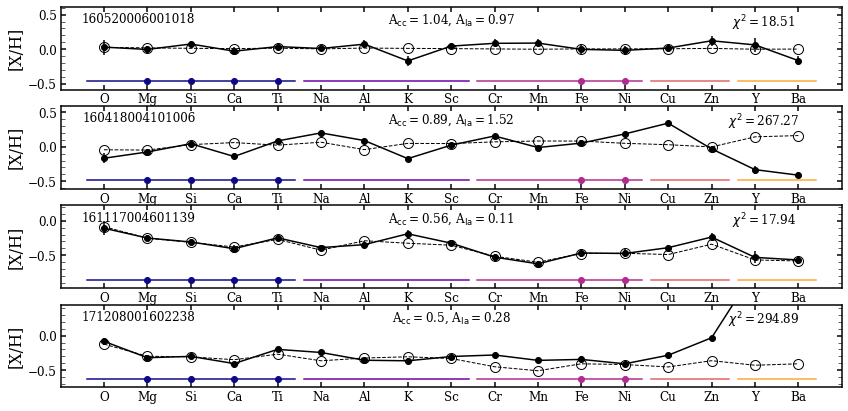}
    \caption{The two-process model predicted (open circles) and reported (solid black circles) [X/H] abundances for four stars in GALAH. The reported abundance points include error bars, though many are smaller than the points themselves. The top two are high-Ia stars near solar [Mg/H]. The bottom two are low-Ia stars with [Mg/H]$<-0.2$. The first and third have $\chi^2$ values near the 50th percentile and the second and fourth have $\chi^2$ values near the 99th percentile, where the $\chi^2$ value is calculated using all elements but Y and Ba. We include the GALAH stellar ID, $\Acc$, $\AIa$ and the $\chi^2$ value for each star in each panel. Colored bars group elements with common physical characteristics. The elements included in the two-process fit are indicated with a colored dot.}
    \label{fig:exp_devs}
\end{figure*}

With the process vectors for all elements and process amplitudes for all stars, we use Equation~\ref{eq:xh} to find the [X/H] values predicted by the two-process model. We do not expect the predictions to perfectly reproduce the abundances of individual stars, in part because of observational errors but also because the model does not account for enrichment mechanisms beyond CCSN and SNIa or for stochastic fluctuations about IMF-averaged yields. As examples of the two-process model predictions and their agreement or disagreement with the GALAH abundances, we plot the observed and predicted [X/H] for four stars in Figure~\ref{fig:exp_devs}. The first two stars have [Mg/H]$\approx 0$ ($\Acc$ near 1) and are in the high-Ia population ($\AIa/\Acc$ near 1). The third and fourth stars have lower metallicity ([Mg/H]$<-0.2$ and low $\Acc$) and are in the low-Ia population (low $\AIa/\Acc$). For each pair of high-Ia and low-Ia stars, we include one with a $\chi^2$ near the 50th percentile (first and third rows, $\chi^2\approx  18$) and one with a $\chi^2$ near the 99th percentile (second and fourth rows, $\chi^2 \approx 255$). The $\chi^2$ value is the sum of the squared differences between the observed and predicted abundances in error units for all elements but Y and Ba. It measures the “goodness” of the two-process model fit. 

Since the first and second stars in Figure~\ref{fig:exp_devs} have $\Acc$ and $\AIa$ near 1, the two-process model predicts values of [X/H]$\approx 0$ for all elements. The first star is well fit by the two-process model, with a $\chi^2$ value of 18.5. We see overlap between the observed and predicted abundances for most elements. The observed Cr, Mn, and Zn values are $\sim0.1$ dex higher than predicted, and the K and Ba values are $\sim0.2$ dex lower than predicted. The second example star deviates much more than the first, with the two-process model correctly predicting only 6 of 17 elemental abundances. Notably, the model overpredicts Cu by $\sim0.4$ dex and underpredicts Y and Ba by $> 0.5$ dex. The low $\chi^2$ and high $\chi^2$ low-Ia stars (third and fourth rows) show similarly good and bad fits, respectively, to the high-Ia examples. The third star's predicted abundances agree with observations and the fourths star's show significant deviations, especially in the Fe-peak, Fe-cliff, and neutron-capture elements. 

We expect that some of the observed deviations from the two-process model predictions indicate problematic spectra or artifacts of faulty data reduction/flagging, but that others identify real enhancements or depletions in the stellar abundances relative to the two-process model. These real differences inform us about the additional non-CCSN and SNIa nucleosynthetic sources and help us identify chemically interesting stars. We will refer to the differences between observed and predicted [X/H] as either ``deviations'' or ``residual abundances''. While these terms are somewhat interchangeable, the second emphasizes our expectation that the two-process description is only approximate, so characterizing a star by $\Acc$, $\AIa$, and the fit residuals is a way to capture major trends with two parameters and focus attention on the (usually small) departures from these trends.

\begin{figure*}[!htb]
    \centering
    \includegraphics[width=\textwidth]{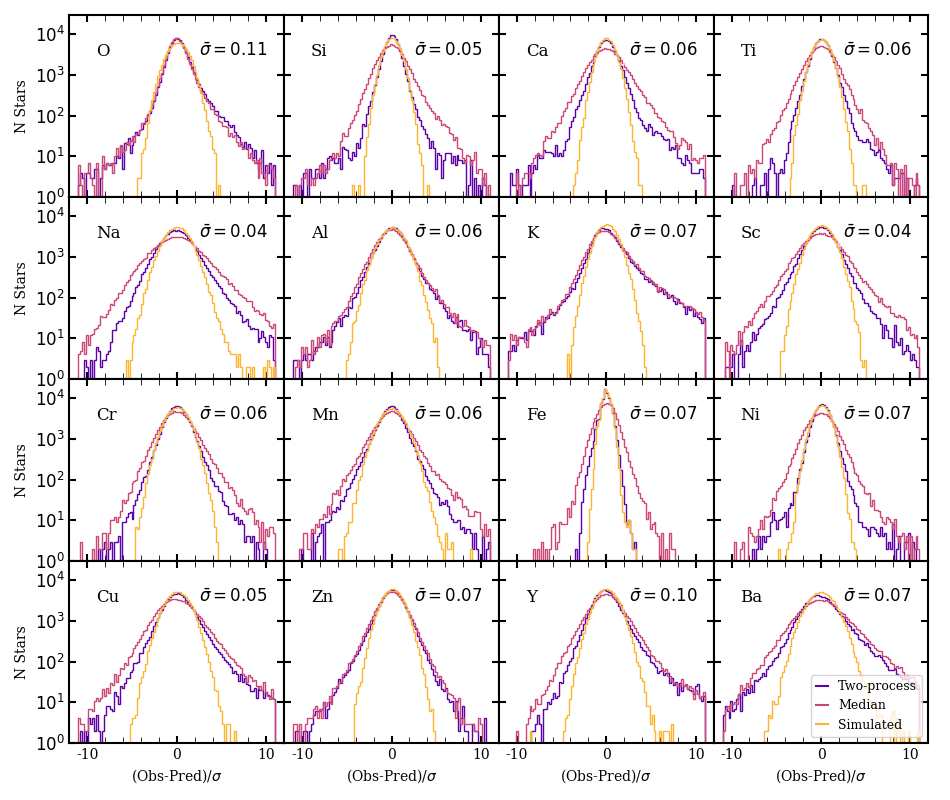}
    \caption{Elemental distributions of residuals from the two-process model. Residuals are calculated as the observed abundance minus the two-process predicted abundance divided by the observational error. We show the distribution of residuals from the two-process model in purple and from the median trends in pink. As a reference for the residuals expected from random Gaussian errors, we plot the distribution of residuals from a population where the abundances are identical to the two-process predictions, plus a random error (orange). The mean $\sigma$ for each element is given in the top right corner of the panel.}
    \label{fig:dev_dist}
\end{figure*}

Before drawing conclusions from the two-process model residuals, we must understand the abundance systematics in our data and establish that the two-process model is a generally good predictor of stellar abundances. In Figure~\ref{fig:dev_dist}, we plot the distribution of [X/H] residuals in error units from the two-process model prediction (purple) and the median [X/Mg] trends (pink), similar to Figure 12 of \citetalias{weinberg2021}. Deviations from the median sequence are found by taking the difference of the observed abundances and the interpolated median value of [X/Mg] at the stellar [Mg/H] for the high-Ia or low-Ia medians. Positive deviations indicate larger observed [X/H] abundances than predicted by the medians and/or two-process model. For all elements, the distribution of two-process residuals is of comparable or smaller width than the distributions of residuals from the medians. This implies that the two-process model predicts a star's abundances more accurately than the median trends of stars with the same [Mg/H] in the same population, in agreement with \citetalias{weinberg2021}'s findings from APOGEE.

At least some of the deviations in Figure~\ref{fig:dev_dist} are an inevitable consequence of observational measurement errors, and in general the two-process model outperforms the median abundance prediction for the elements with the smallest observational uncertainties. However, for nearly all elements there are stars with $>8\sigma$ deviations from the two-process predictions, indicating either true residuals that are large compared to the observational uncertainties or non-Gaussian tails on the observational error distribution, or both. To quantify the expected distribution of residuals from observational error alone, we construct a population with ``simulated'' abundances. Stars in this population adopt the stellar [X/H] abundances predicted by the two-process model with the star's best-fit $\Acc$ and $\AIa$. We then add a random error from a Gaussian distribution with $\sigma$ equal to the reported error on the stellar [X/Fe] abundances, representing the observational noise. This population represents a sample that the two-process model could have perfectly predicted in the absence of noise. We plot the resulting distribution of residuals for the simulated population in Figure~\ref{fig:dev_dist} as a prediction of what the distributions would look like if only Gaussian observational noise were present.

For all elements, the core of the simulated distribution closely resembles the core of the distribution of two-process residuals but the wings of the observed distribution are much wider, with clear differences setting in beyond $3-4\sigma$. In themselves, these distributions do not tell us whether many-$\sigma$ residuals arise from true deviations from the two-process model or from observational errors that are large compared to a Gaussian distribution with the reported $\sigma$. Our analysis below will demonstrate examples of both. We note that the agreement in the cores of the distributions suggests that GALAH's reported abundance uncertainties are accurate (or possibly overestimated) for most stars for all of these elements. 

In Figure~\ref{fig:dev_dist}, we see that not all of the two-process deviation distributions are symmetrical. The distributions of O, Na, Al, K, Cu, Y, and Ba are skewed such that there are more stars with excesses of these elements than depletions relative to the two-process model. These asymmetries could be a sign of correlated residuals, e.g., populations of stars in which an additional enrichment process or a stochastic variation in CCSN or SNIa yields causes extra production of multiple elements. 

Alternatively, the asymmetry could indicate systematic biases in the data reduction. In particular, for O and K departures from LTE are significant. Although parameter-dependent non-LTE corrections are implemented for GALAH DR3 to mitigate this effect (typically decreasing the measured abundance), uncertainties in the stellar parameters can propagate through to uncertain non-LTE corrections. For K, higher abundances can also be caused by absorption features from interstellar K contaminating the spectrum. For Na, Al, and Cu, we have no obvious observational explanation for this particular skewness. For Y and Ba, both measured via ionised lines, higher values might be caused by uncertainties in the surface gravities, which influence these lines much more than neutral lines. However, both of these elements are expected to have large contributions from AGB enrichment, so physical departures from the two-process predictions would not be surprising.

\subsection{Correlated Residuals} \label{subsec:cors}

As emphasized by \citet{ting2021}, the correlations or covariance of residuals can demonstrate the reality of intrinsic abundance fluctuations even when the typical residuals for an individual star are comparable to the observational uncertainty. For two elements whose residual abundances are correlated with correlation coefficient $\rho$, the statistical uncertainty in a sample of $N$ stars is $\Delta \rho \approx 1 / \sqrt{N}$, so relatively small correlations can be measured at high statistical significance in a large sample. Most sources of observational uncertainty produce errors that are nearly uncorrelated from one element to another, so a non-zero covariance measurement can provide evidence of an intrinsic correlation even if the exact magnitude of observational uncertainties is not perfectly known. (We discuss one important caveat to this statement below.) Furthermore, correlations provide insight on the possible physical origin of residual abundance fluctuations, in addition to their magnitude.

We compute the covariance of pairs of elements
\begin{equation}\label{eq:cov}
    C_{ij} = \langle\Delta_i \Delta_j\rangle
\end{equation}
with 
\begin{equation}\label{eq:cov_i}
    \Delta_i \equiv [\textrm{X}_i/\textrm{H}]_{\text{obs}} - [\textrm{X}_i/\textrm{H}]_{\text{pred}},
\end{equation}
where the predicted values are the abundances derived from the two-process model fit with 6 elements. We remove all elemental deviations $>10\sigma$ from the covariance calculations, as these outliers are likely due to reduction errors and were found to drive some correlations in \citetalias{weinberg2021}. The removal has a small effect on the neutron-capture elements, reversing the sign of the Ba-Ti and Y-Ca covariances and strengthening the Ba-Cu negative covariance. Changing the cut from $10\sigma$ to $5\sigma$ has little further effect, indicating that the covariances we measure are not driven by stars in the extreme tails of the residual abundance distributions. 

\begin{figure*}[!htb]
    \centering
    \includegraphics[width=0.8\textwidth]{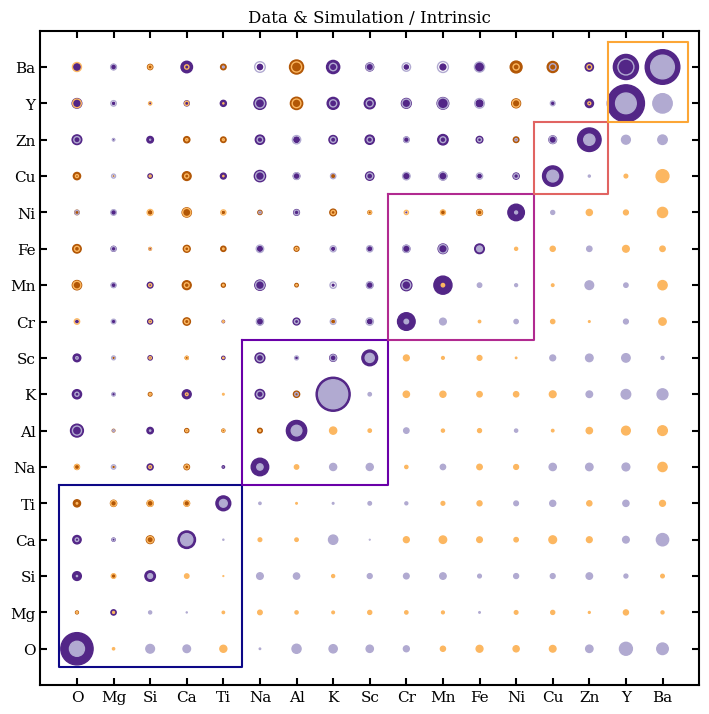}
    \caption{\textit{Above Diagonal}: The covariance matrix of the residuals between the observed and two-process predicted abundances. Positive values are shown in dark purple and negative values in orange. The area of the circle scales with the magnitude of the covariance. For visual scaling, the diagonal elements (O, O) and (Ba, Ba) have values of about (0.02)$^2$, (Zn, Zn) has a value of about (0.01)$^2$, and (Na, Na) has a value of about (0.005)$^2$. We plot the covariance of the simulated data set, the covariance expected from random Gaussian errors, as open circles, which usually appear as light purple or light orange circles within the larger dark circles that represent the observed covariance. We place squares around the $\alpha$ (blue), light odd-$Z$ (purple), Fe-peak (pink), Fe-cliff (orange) and neutron-capture (yellow) elements to guide the eye. \textit{Along Diagonal}: The area of the circle represents the total variance of the residual abundance, and the area of the light-shaded circle represents the inferred intrinsic variance (data $-$ simulation), which is slightly negative (orange) in a few cases. \textit{Below Diagonal}: We plot the intrinsic covariance matrix, (data - simulation) as solid light purple circles (positive values) and solid light orange circles (negative values).}
    \label{fig:covs}
\end{figure*}

We plot the covariances of residual abundances for our 17 elements along and above the diagonal in the Figure~\ref{fig:covs}, where dark purple circles represent positive covariances and dark orange circles represent negative covariances. The magnitude of the covariance scales with the area of the circle, such that the diagonal elements (O, O) and (Ba, Ba) have values of about (0.02)$^2$, (Zn, Zn) has a value of about (0.01)$^2$, and (Na, Na) has a value of about (0.005)$^2$. In this matrix, the diagonal elements are the square of the RMS deviations, so we see larger diagonal elements for those with larger scatter (O, K, Y, Ba) and smaller diagonal elements for those with low scatter (Mg, Si, Ti, Fe, Ni). Off the diagonals we see how the residuals for one element correlate with the residuals for another. If the covariance between X and Y is positive, stars that have a higher abundance of X than predicted by the two-process model are likely to have a higher abundances of Y. Strong positive covariances between a set of elements may indicate that those elements have an additional enrichment source not included in the two-process model. 

Above the diagonal in Figure~\ref{fig:covs} we show the covariance of the data (dark tone, solid circles) and the covariance of the simulated data set (light tone, open circles, see Section~\ref{sec:twoproc}). The simulated covariance indicates the level of covariance we expect from the observational errors alone, assuming that the errors themselves are uncorrelated and Gaussian with the reported RMS scatter. As discussed in detail by \citet{ting2021} and \citetalias{weinberg2021}, correlated residuals still arise in this case because the values of $\Acc$ and $\AIa$ fluctuate around their true values, and a random error in these parameters leads to correlated deviations among multiple elements (Equation 47 of \citetalias{weinberg2021}). Following \citet{ting2021}, we refer to this artificially inferred covariance as ``measurement aberration,'' which arises from computing abundance residuals with respect to an imperfect reference. In most cases, the measured covariance exceeds the simulated covariance, implying a true intrinsic covariance of the same sign but somewhat reduced magnitude. In a few cases (e.g., Ba-Ca) the simulated covariance is opposite in sign to the measured covariance, implying an intrinsic covariance that is still larger. We plot the inferred intrinsic covariance matrix (data - simulation) along and below the diagonal in Figure~\ref{fig:covs} as the solid, light toned circles.

Using APOGEE data, \citetalias{weinberg2021} identify two groups of elements with two-processes residuals that positively correlate, one comprised of Ca, Na, Al, K, and Cr, and the other comprised of Ni, V, Mn, and Co. Correlated patterns are somewhat hard to pick out of Figure~\ref{fig:covs}, perhaps because the GALAH abundance errors are slightly larger than the APOGEE abundance errors and measurement aberration is therefore a larger relative effect. Nonetheless, we note the following trends.
\begin{enumerate}
    \itemsep0em 
    \item All of the Fe-peak elements have positive intrinsic covariance with each other except Ni and Cr with Fe, though the simulated covariance is comparable to that of the data in many cases. The Fe-Ni anti-correlation probably arises because both elements are used in the two-process fit. 
    \item Ca and K residuals have a negative covariance with all of the Fe-peak elements. 
    \item The Ba and Y residuals are positively correlated with each other and with Zn residuals. 
    \item Al and Ni have a clear anti-correlation with Zn, Ba, and Y.
    \item Si residuals have a positive intrinsic covariance with all light odd-$Z$ and Fe-peak elements but K. 
\end{enumerate}

The covariances of the Cu, Zn, Ba, and Y residuals are most interesting, as these elements are expected to have contribution from AGB stars, and Ba and Y are expected to have little contribution from SNIa. The two-process model attributes all delayed nucleosynthesis to SNIa, which may poorly describe the abundances of stars with significant AGB contribution. We would thus expect to see a correlation in the abundance residuals for elements with an AGB component. 

\subsection{Correlations with Age}\label{subsec:age}

\begin{figure}[!htb]
    \centering
    \includegraphics[width=\columnwidth]{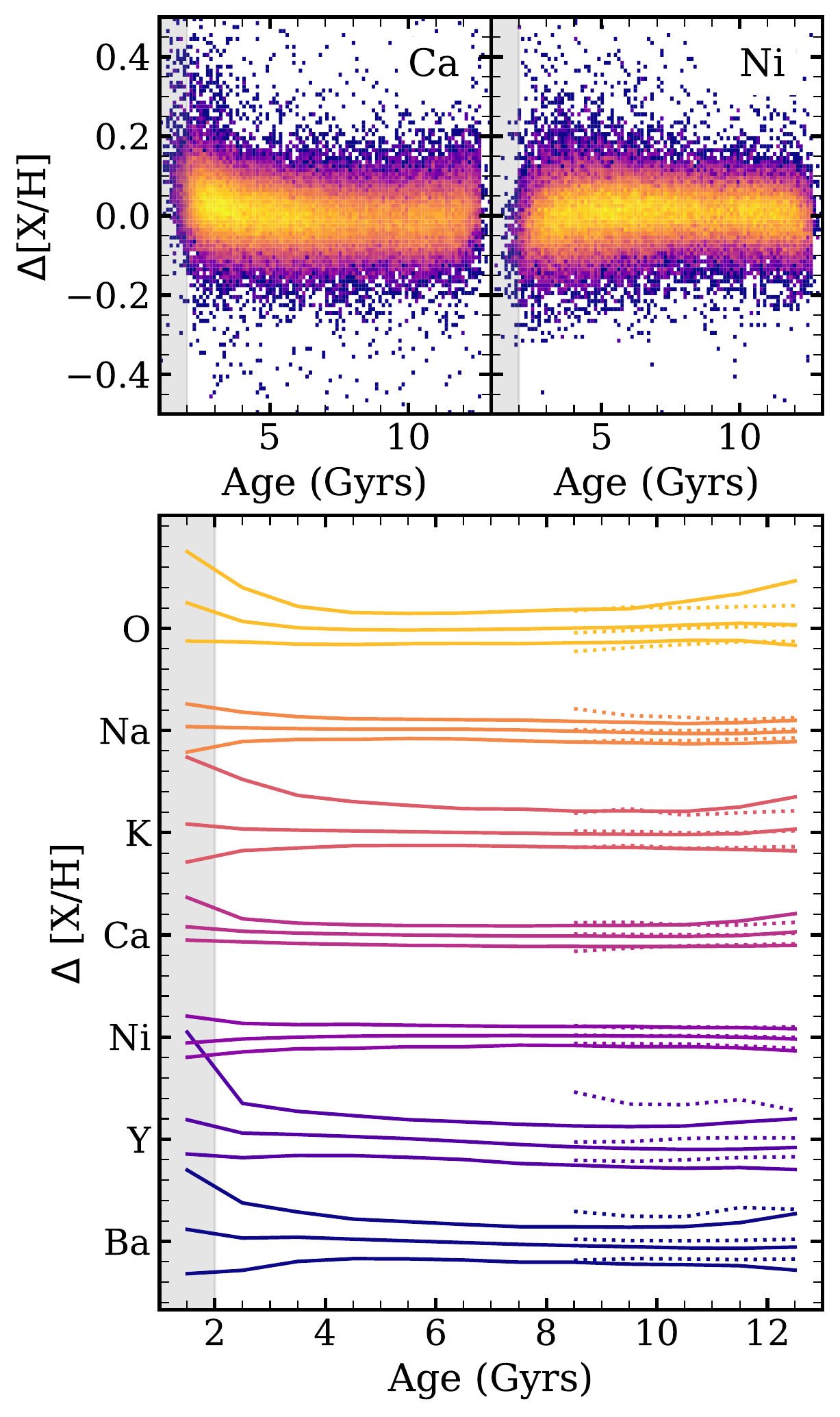}
    \caption{Correlation between stellar residual abundances and ages for seven elements. \textit{Top}: Distribution of $\Delta$[X/H] vs. age for Ca (left) and Ni (right), colored by number density. \textit{Bottom}: 5th, 50th, and 95th percentile contours of the $\Delta$[X/H] vs. age distribution for the high-Ia (solid lines) and low-Ia (dotted lines) populations. We plot the trends for O, Na, K, Ca, Ni, Y, and Ba from top to bottom. The separation of small tick marks is 0.2 dex, and successive elements are displaced by 1 dex. We shade the region from 0 to 2 Gyrs grey to indicate uncertain ages, as those below and near 1 Gyr may be overpredicted or missing because of our sample selection (Section~\ref{subsec:sample})}
    \label{fig:delta_age}
\end{figure}

In Figure~\ref{fig:delta_age}, we show the correlation of $\Delta$[X/H] with age for seven elements. We take BSTEP ages \citep{sharma2018} from the GALAH value-added catalog for stars with fractional errors $<25\%$, as in Section~\ref{subsec:As}. In the top panel we plot density maps of $\Delta$[X/H] vs. age for Ca and Ni. For both elements the core of the distribution is at 0, but there are stars with larger residuals at all ages. We see a small positive upturn in the tail of the Ca residuals at young ages.

To better understand the elemental residual-age trend of the core and the tails, we plot the 5th, 50th, and 95th percentile contours of the distribution for O, Na, K, Ca, Ni, Y, and Ba in the lower panel of Figure~\ref{fig:delta_age}. We separate the high-Ia (solid lines) and low-Ia (dotted lines) populations for clarity. As is seen in the top panels, there are no trends in the residual abundances with age for Ni and an upturn in the 95th percentile contour for Ca at young ages ($<$3 Gyrs). Like Ca and Ni, Na does not show strong trends with age, though the tails of the residual abundances ``flare'' at young ages. The Cu, Ti, and Sc trends resemble that of Ca, and Zn trends resemble Na. Fe, Mn, Cr, Si, Al, and Mg residuals (not illustrated) show no correlation with age. The lack of residual abundance correlation with age for these elements suggests that the age-dependent enrichment (e.g., SNIa) has been properly accounted for by the two-process model.

We observe stronger correlations between the stellar abundances residuals and ages for K, O, Y, and Ba. These four elements have the larger RMS deviations (Figure~\ref{fig:covs}) and are skewed to positive $\Delta$[X/H] in Figure~\ref{fig:dev_dist}. The K residual-age correlation is unique, as the core stays at zero but there is considerable flaring to low (0.3 dex) and high (0.8 dex) deviations in the youngest stars. The trends of O, Y, and Ba resemble each other, as the 50th and 95th percentiles are inclined for stars younger than 4 Gyrs. The 0.1-0.2 dex rise in the median that younger stars have higher Ba, Y, and O enrichment than predicted by the two-process model. This is surprising for O, a pure CCSN element with no known time dependent enrichment source, but unsurprising for neutron-capture elements Y and Ba. Both elements have delayed AGB enrichment that is only indirectly accounted for in the two-process model. This correlation is in agreement with the observed enhancement of Y and Ba in young open clusters \citep[][see Section~\ref{sec:clusters}]{spina2021, baratella2021, casamiquela2021} and the residual abundance-age correlation of Ce, another neutron-capture element with AGB enrichment, in \citetalias{weinberg2021}.

However, as noted in Section~\ref{subsec:As}, our sample includes a population of $\sim$6000 stars that are rotationaly broadened ($v_{\rm broad}>20$ km/s). The rapid rotaters are young (age $< 4$ Gyrs) hot ($\teff > 6000$ K) stars \citep[e.g.,][]{vansaders2013} whose rotational broadening hampers accurate abundance determination and skews the two-process fit to high values of $\AIa/\Acc$. The exclusion of stars with $v_{\rm broad}>20$ km/s does not affect the trends described above, but stricter cuts reduce the strength of abundance residual-age correlations observed in Figure~\ref{fig:delta_age} for O, K, Y, and Ba. Further investigation into the impact of rotational broadening on the abundance determination of young stars is necessary to fully understand the residual abundance correlations with age.

\subsection{What Fraction of the Large Deviations are Real?} \label{subsec:real?}

\begin{figure}[!htb]
    \centering
    \includegraphics[width=\columnwidth]{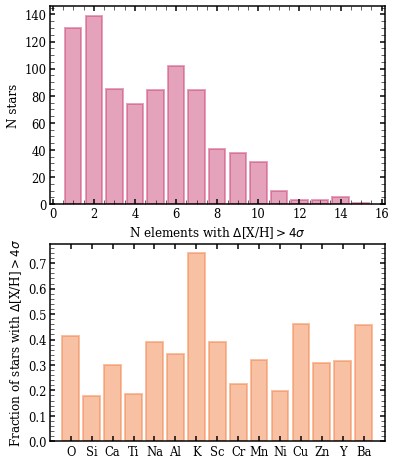}
    \caption{Statistics of stars in the 99th percentile of the $\chi^2$ distribution. \textit{Top:} Number of these stars with $N$ elements showing deviations greater than $4\sigma$. \textit{Bottom:} Fraction of these stars with $\Delta \xh > 4\sigma$ for each element.}
    \label{fig:99stat}
\end{figure}

To better understand the stars with large deviations, we take a closer look at those in the 99th percentile of the $\chi^2$ distribution. We select the elements for each star that have a deviation greater than four times their reported abundance error ($|\Delta \xh| > 4\sigma$), as a diagnostic of the number of highly deviating elements. Figure~\ref{fig:99stat} plots the number of 99th percentile stars that have $N$ elements with $|\Delta \xh| >4\sigma$ in the top panel and the fraction of these stars that have $|\Delta \xh| >4\sigma$ for each element in the bottom panel. Of the 830 stars in the 99th percentile of the $\chi^2$ distribution, 169 have one or two elements with predicted abundances that deviate from the observed abundances by more than $4\sigma$ and 345 stars have 3-6 highly deviating elements. The rest of the stars have 7 or more elements with deviations greater than 4$\sigma$. Of all the elements, K exhibits strong deviations the most frequently, with 74\% of 99th percentile stars displaying K deviations over $4\sigma$. The high fraction of stars with significant K deviations could plausibly arise from uncertainties in the NLTE corrections or from interstellar contamination. Si, Ti, and Ni are well measured in GALAH and show thinner, symmetrical deviation distributions in Figure~\ref{fig:dev_dist}. These elements are the least represented in the high-$\chi^2$ population, with $>4\sigma$ deviations arising in less than $20\%$ of these stars. 

\citetalias{weinberg2021} studied a sample of high $\chi^2$ stars in APOGEE and found a mix of stars with peculiar abundances and unflagged quirks that affect the measurements (such as high rotational velocities, spectroscopic binaries, and absorption lines falling on chip gaps). Quantifying the relative number of true physical outliers vs. large observational errors requires careful, systematic investigation of a representative sample of high-$\chi^2$ stars. To this end, we inspected a set of 100 stars in the 99th percentile of the $\chi^2$ distribution. We plotted the SME output for all unflagged line windows for all 100 stars, paying close attention to the elements that significantly deviate from the two-process model predictions. We checked that the distribution of the number of stars with $N$ highly deviating elements and the distribution of deviating elements resemble those of the full population (Figure~\ref{fig:99stat}). We classified each star's SME fit as good (all highly deviating lines are well fit), bad (all highly deviating lines are poorly fit), or okay (some highly deviating lines are well fit and others are poorly fit), noting double peaked lines, asymmetric lines, emission features, and poor wavelength solutions.

\begin{figure*}[!htb]
    \centering
    \includegraphics[width=\textwidth]{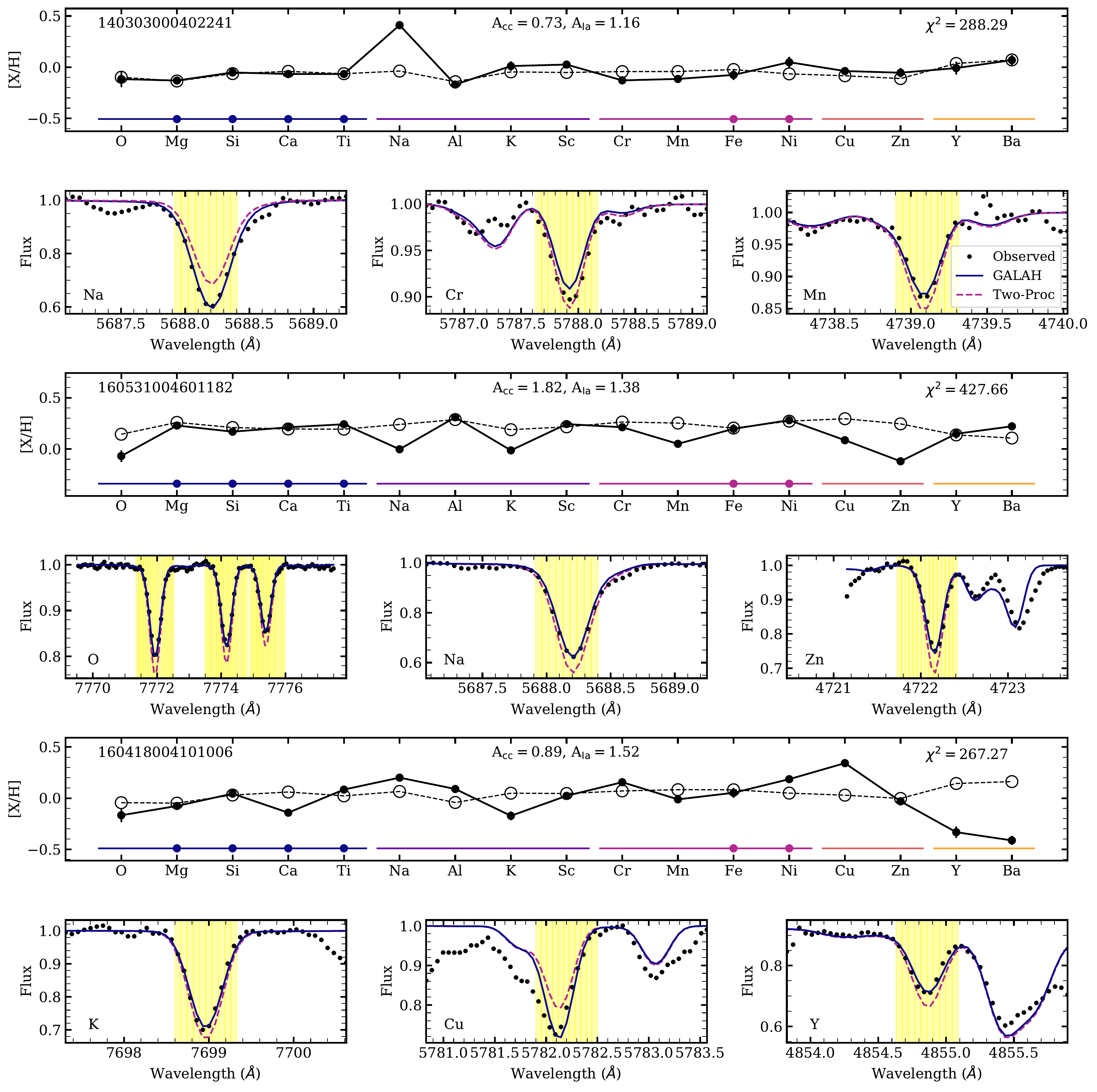}
    \caption{Observed and predicted abundances for three stars in the 99th percentile of the $\chi^2$ distribution. The first, third, and fifth rows are the same as Figure~\ref{fig:exp_devs}. The second, fourth, and sixth rows plot the line windows for three elements, listed in bottom left hand corner of each panel, for the star in the row above. In each line window we plot the observed, normalized flux as black points, the GALAH best-fit line profile as a dark blue line, and the line profile inferred from the two-process abundances as a pink dashed line. The wavelength window fit by SME is shaded yellow.}
    \label{fig:lines}
\end{figure*}

From this analysis, we identify a sample of stars that have well fit spectra and high $\chi^2$ relative to the two-process prediction. We plot the observed and two-process predicted abundances for three of these stars in Figure~\ref{fig:lines}, including spectra of three line windows for elements that show deviations from the two-process model predictions for each star. We plot the observed, normalized flux as points with error bars (though the error bars are smaller than the points in all cases) and the GALAH SME fit as solid dark blue lines. With SME, we are also able to construct the line profiles for each star if it had the two-process predicted abundances. For each element line in question, we alter the absolute abundance of that element in the SME abundance structure, but we otherwise use the same SME setup as the outcome of the final GALAH DR3 fit. This comparison allows us to see how discrepant the two-process model abundances are with the observed lines and helps us to understand which enhancements and depletions are significant. In Figure~\ref{fig:lines}, the two-process predicted line profiles are plotted in pink and the light yellow shaded region indicates the line window used in the SME fit. We describe the three stars below. 

140303000402241: For this star, the two-process model predicts abundances near the observed values for all elements but Na, which deviates by $>4\sigma$. We see that the $\alpha$-elements are all well fit and that the light odd-$Z$ and Fe-peak elements have deviations of $\sim0.1$ dex. We show a Na, Cr, and Mn line window for this star. For Na and Mn, the GALAH SME lines fit the observed data better than the those predicted by the two-process fit. While we only show one Na line here, the second observed Na line at 5682.6\r{A} is also deeper than line predicted by the two-process model abundance. Conversely, the Cr window shown here appears better fit by the two-process abundance than the GALAH abundance--though neither line passes through the data points at the line's peak. Of the three strong Cr lines, one (shown here) is better fit by the two-process predicted abundance, one by the reported GALAH abundance, and one sits halfway between the GALAH and two-process values. Since GALAH is fitting all lines simultaneously, the fits to individual lines may be poor if they independently suggest different abundances. In this star, the small Cr deviation indicates abundance uncertainty, but the $>4\sigma$ deviation in Na and the smaller Mn deviation are real.

160531004601182: The second example star has a larger $\chi^2$ value than the first and shows significant deviations in six elements. O, Na, K, Mn, Cu and Zn are all reported as 0.2 to 0.4 dex lower than the two-process model predictions, with all but O deviating by more than $4\sigma$. We show the O triplet, one of two Na lines, and one of two Zn lines. In all cases the two-process abundances overpredict the depth of all lines for these elements. This star has real depletion in O, Na, and Zn relative to the two-process model predictions.

160418004101006: Our final star shows a mix of positive and negative deviations from the two-process model, with $|\Delta \xh| > 4\sigma$ for Ca, Na, K, Cu, Y and Ba. We show a K and Y line, both overpredicted by the two-process model, and a Cu line, underpredicted by the two-process model. As for the second star, all GALAH SME lines fit the observed data better than the lines inferred from the two-process model, indicating real deviations. 

For all three stars we find that the highly deviating lines ($|\Delta \xh| > 4\sigma$) are well fit by GALAH and are inconsistent with the two-process predicted abundances. However, of the 100 inspected stars, we only find seven with abundance deviations that that are convincingly real as in these three examples. Each of these star's abundances are interesting and could indicate unique chemical enrichment histories. We discuss the population of high-Na stars (like the first star of Figure~\ref{fig:lines}) later in this section.

The spectra of the other 93 stars show highly deviating lines that are poorly fit by the GALAH analysis. We identify 60 as having bad fits, where the spectral features indicate that none of the highly deviating lines (or often any lines) should be trusted. At least one third of these stars (22) exhibit spectral signatures of binarity, such as double peaked O lines, broad features, and/or asymmetric lines. The GALAH DR3 pipeline uses several algorithms to automatically identify binaries \citep{buder2021}. This includes cool main sequence stars that are significantly more luminous than can be explained with the most luminous isochrones, which can however only be applied up to a certain $\teff$, before the turn-off stars overlap with the binary main sequence. The second algorithm uses the spectral classification algorithm tSNE \citep{traven2017} to identify line-split binaries. The classification algorithm fails, however, to distinguish between fast rotating stars and binaries, when the lines are broadened, but not split. The latter stars are more likely to go undetected and end up in our sample. Eight of the 60 stars with bad fits have poor wavelength solutions, and 14 show emission features, often in the K, O, and Al windows.

The remaining 33 stars exhibit some highly deviating elements with well fit lines but some with poor fits. We inspected the lines of all elements with $|\Delta \xh| > 4\sigma$, a total of 148 elements across 33 stars, excluding Y and Ba. $48\%$ of the highly deviating elements are well fit in all of their windows, indicating real deviations. The remaining $52\%$ of elements are poorly fit in at least one line window and suffer from observational errors, emission lines resulting from poor telluric subtraction (10 stars), or poor SME fits. Only spectral inspection can identify which elements' deviations are real and which are untrustworthy. 

Based on our analysis of this small sample, we estimate that roughly 40\% of the stars near the 99th percentile of the $\chi^2$ distribution have real deviations, but many of these stars have a mix of genuine large abundance residuals and incorrect measurements. The remaining 60\% have problematic data that affect many elements, with no convincing genuine deviations. Roughly 45\% of the stars ``should'' be flagged but are not: 20\% for binarity, 10\% for poor wavelength solutions, and 25\% for poor telluric subtraction of one or more lines. 

Of these 100 inspected stars, 18 have a single highly deviating element, with 17 showing emission features or an asymmetric line. In 13 cases, the lone deviating element is K. It is unsurprising that selecting the top one-percent of deviating stars identifies many cases with unusual measurement errors. The high fraction of measurement systematics in this sample does not imply that similar systematics affect most GALAH stars, nor does it imply that less extreme deviations from the two-process model are typically caused by measurement systematics. As abundance pipelines improve with successive data releases (which has certainly been the case for APOGEE), we expect that a larger fraction of measurement problems will be corrected or at least flagged, so that a larger fraction of the high-$\chi^2$ stars are truly chemically peculiar.

\begin{figure*}[!htb]
    \centering
    \includegraphics[width=\textwidth]{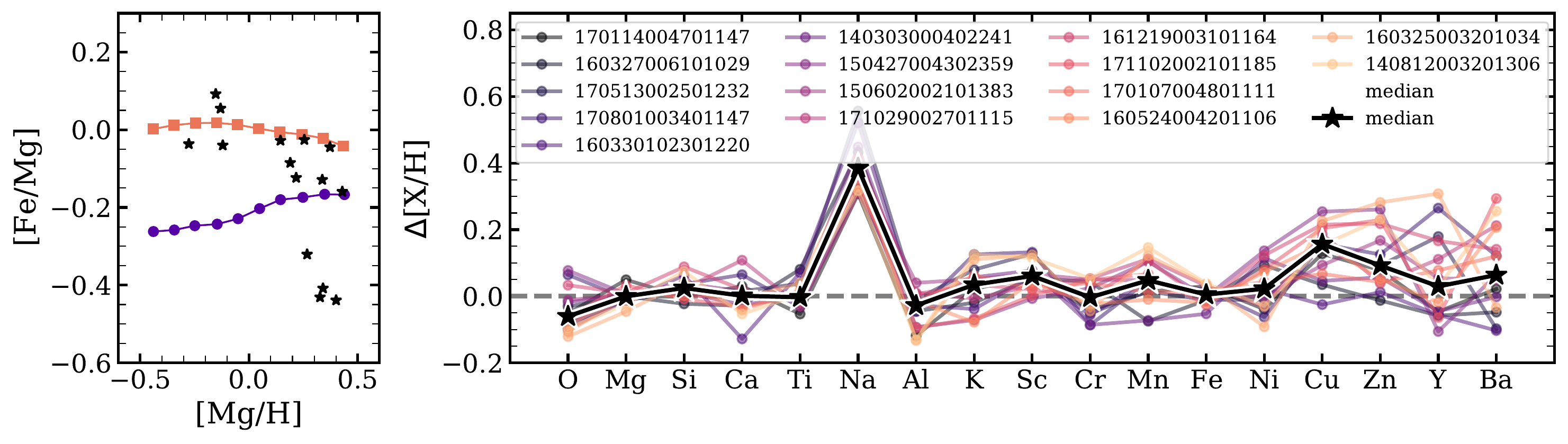}
    \caption{Abundance deviation from the two process model (observed - predicted) for 15 stars with high Na deviations ($>0.3$ dex) and low ($<0.15$ dex) deviations of all other elements, O through Ni. We plot each star as its own line, listing the GALAH object IDs in the legend. We plot the median deviation of these 15 stars as the solid black stars. For context, the left panel shows the location of these stars in [Fe/Mg] vs. [Mg/H] relative to the low-Ia and high-Ia median sequences.}
    \label{fig:Na}
\end{figure*}

Residual abundance analysis presents the opportunity to find populations of stars whose abundances are exceptional relative to other stars with similar levels of CCSN and SNIa enrichment. While chemically peculiar stars are not the focus of this paper, we illustrate this opportunity with the example of stars like the first star of Figure~\ref{fig:lines}, with a strong excess of Na. We find a total of 15 stars that have most elemental abundances close to the two-process predictions (O to Ni within 0.15 dex) and a Na deviation $>0.3$ dex. Inspection of the stellar spectra does not identify any obvious issues, although one of the stars has broad lines. We plot the deviation from the two-process abundance ($\Delta \xh$) for the 15 high Na stars in Figure~\ref{fig:Na} along with the median deviation of all 15. Seven of these stars have Cu, Zn, Y, and/or Ba deviations in excess of $0.2$ dex, leading to median $\Delta$[Cu/H] of 0.16 dex and a median $\Delta$[Zn/H] of 0.09 dex. These common deviations could indicate a rare astrophysical sources that efficiently produces all of these elements. We find no immediate evidence of a similarity in Galactic location, eccentricity, stellar parameters, age, or orbital dynamics for these 15 stars. Intriguingly, four of them have unusually low values of [Fe/Mg], though overall they are widely spread in the [Fe/Mg] vs. [Mg/H] diagram.

Correlations of residual abundances and rarer large deviations from two-process predictions hold a wealth of information about nucleosynthesis and Galactic enrichment history. In future work we will conduct a more comprehensive search for populations with like deviations and strive to understand what may cause enhancements and depletion among these stars.

\section{Residual Abundances of Open Clusters} \label{sec:clusters}

Open clusters, groups of stars that form from the same gas at the same point in our Galactic history, are expected to have uniform stellar abundances. If a cluster is enhanced in some element, we expect that all stars in that cluster will show similar enhancements. While processes such as atomic diffusion, planet formation, and planet engulfment can cause surface abundance variations between co-natal stars \citep[][and references therein]{casamiquela2021}, many works have measured the level of homogeneity among cluster members to be within $0.02-0.03$ dex \citep{desilva2006, liu2016b, bovy2016, casamiquela2020, ness2021}. In this section we study the residual abundances of known open clusters in GALAH membership and cluster age taken from \citet{spina2021}. With multiple stars per cluster, we use median abundances to reduce statistical uncertainties and the impact of rare systematic errors on residual abundance. The ages of young open clusters derived from color-magnitude diagram fitting may be more accurate than individual stellar isochrone ages, especially for ages $\lesssim 1$ Gyr, providing another avenue to study the residual abundance trends with age, as in Section~\ref{subsec:age}. By studying the clusters' residual abundance trends with age we can also investigate whether or not clusters of the same age have distinct residual abundance patterns. If so, this would be a positive sign for chemical tagging, which strives to leverage stellar abundance similarities to identify co-natal populations \citep[e.g.,][]{freeman2002}. Here we do not attempt to address the question of homogenity within clusters, as that is better done with more targeted data of even higher resolution and SNR.

\begin{figure*}[!htb]
    \centering
    \includegraphics[width=\textwidth]{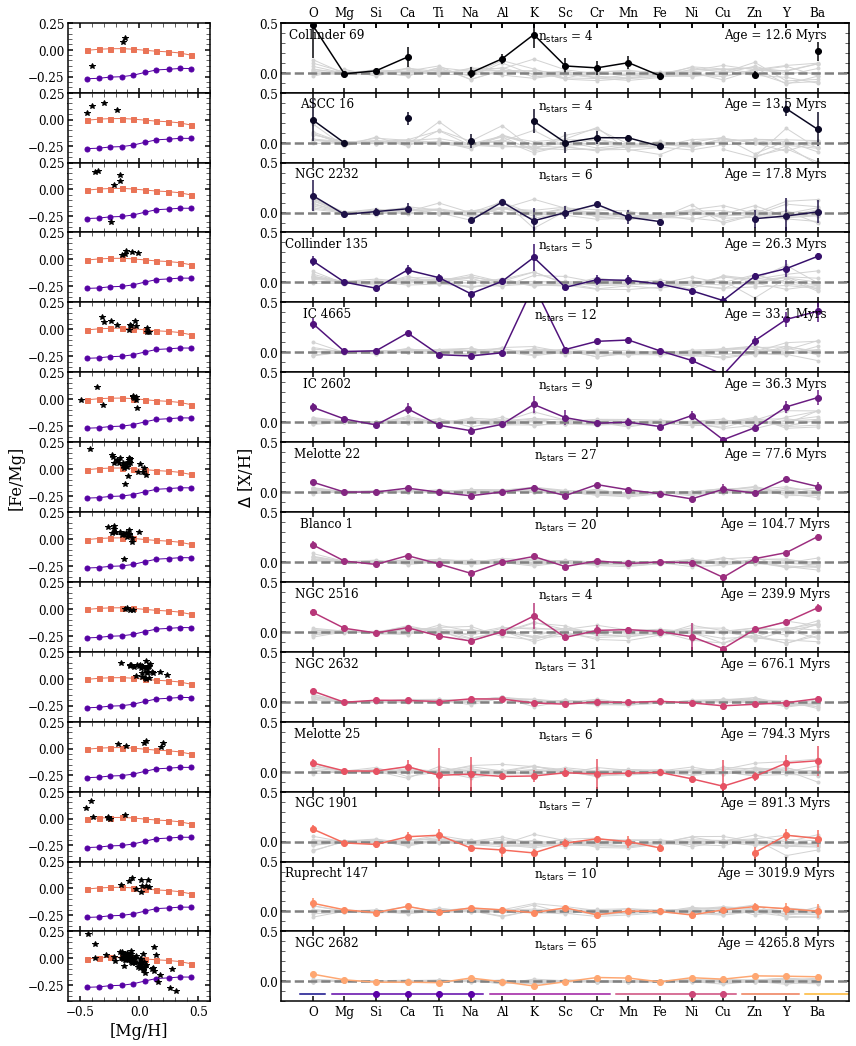}
    \caption{\textit{Left}: [Fe/Mg] vs. [Mg/H] abundances for 14 open clusters. In each panel we show the median high-Ia (orange squares) and low-Ia (purple circles) trends and the abundances of all cluster members (black stars). \textit{Right}: Abundance residuals ($\Delta\xh$, obs-pred) for 14 open clusters. Cluster name, age, and the number of stars (N) included in this analysis are listed in all panels. For each cluster, we plot the median residuals of all elements with $\geq4$ stars, with error bars representing the standard deviation of the median of 1000 bootstrapped cluster samples (colored lines). Background grey lines show the median deviation of 10 field samples of N stars with [Fe/H] and [Mg/Fe] within 0.05 dex of the cluster median. Clusters are plotted in order of increasing age.}
    \label{fig:cluster2}
\end{figure*}

To identify potential cluster members, we cross match our stellar sample with the open cluster catalog from \citet{spina2021}, who identify cluster members and assign membership probability from \textit{Gaia} astrometry. We define stars as cluster members if they have membership probability $>70\%$. This cut leaves us with 14 clusters of four or more stars. The left hand panels of Figure~\ref{fig:cluster2} show the [Fe/Mg] vs. [Mg/H] values of each cluster star with median high-Ia and low-Ia trends for our entire population. While some clusters are concentrated in [Fe/Mg] and [Mg/H] (e.g., Collinder 135), others span a range of over 0.5 dex in [Mg/H] (e.g., NGC 2682). This sizable variation is unexpected given the predicted uniformity of cluster abundances. \citet{spina2021} do not comment on the range of metallicities in their clusters, but this could point to contamination by field stars. Here we take cluster membership at face value, but contamination is a possible limitation of our analysis.

\begin{figure*}[!htb]
    \centering
    \includegraphics[width=\textwidth]{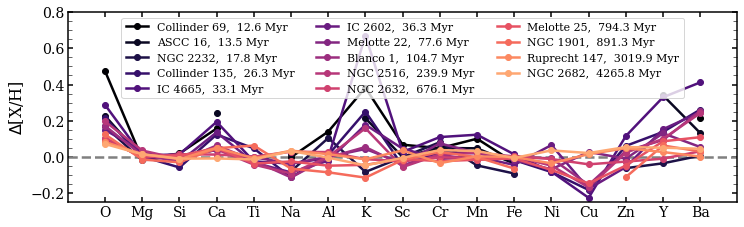}
    \caption{Median deviations of open clusters with $>4$ stars in GALAH. Clusters are color coded by increasing age, with youngest clusters in black and oldest clusters in yellow. Solid lines connect the deviations for each cluster to guide the eye.}
    \label{fig:cluster}
\end{figure*}

In the right panels of Figure~\ref{fig:cluster2} we plot the median $\Delta\xh$ (observed - predicted) for each cluster, excluding median values for elements with less than four unflagged [X/Fe] abundances. Error bars represent the standard deviation on the medians of 1000 bootstrapped samples of each cluster. Because the error on the median scales like $\sigma/\sqrt{N}$, where $N$ is the number of cluster members, the uncertainties are larger for clusters with fewer members.
We also plot 10 example medians of $N$ field stars with [Fe/H] and [Mg/Fe] within 0.05 dex of cluster median. The uncertainty again scales with $N$, such that the median abundance residuals span $\sim 0.1-0.15$ dex in random field samples of 4-6 stars while residuals from larger samples ($N>10$) are smaller, often less than $0.05$ dex. To better compare the median abundance residuals of all clusters, Figure~\ref{fig:cluster} plots the median $\Delta\xh$ for all 14 of them in one panel. Clusters are colored by increasing age, with the youngest clusters in black and the oldest in yellow. 

In Figure~\ref{fig:cluster2}, the median residual abundances for most elements in clusters NGC 2632 and NGC 2682 sit close to zero and within the range of field star deviations. These two clusters are known to have solar abundances \citep[e.g.,][]{boesgaard2013, liu2019}, so their small abundance residuals are not surprising. In other clusters we see more significant variations from the field samples, especially in O, Ca, K, Cu, Y, and Ba. When comparing the median abundance residuals of all clusters we see many overarching trends, most notably that all clusters have positive O residuals and that younger clusters have larger abundance residuals than older clusters. The youngest clusters exhibit higher Ca and lower Cu than predicted by the two-process model. We further observe excesses of K, Y, and Ba, though the K should be interpreted with caution due to known systematics and higher scatter. 

We expect to see the positive Ba and Y residuals, as both elements display supersolar [X/Fe] abundances in \citet{spina2021}. The enhancement of Ba in open clusters was identified in \citet{dorazi2009}, who found strong enrichment of Ba ([Ba/Fe]$\approx 0.6$ dex) that decreases with cluster age. Recently, high [Ba/Fe] and [Y/Fe] has also been observed in young clusters by \citet[][\text{Gaia}-ESO]{baratella2021} and \citet{casamiquela2021} and high [Ce/Fe] as observed in young clusters by Sales-Silva (in prep.). While we tend to see larger [Ba/H] residuals for younger clusters, the age trend is less obvious in our data. \citet{dorazi2009} conclude that such high Ba abundances cannot be produced from standard nucleosynthesis in the young clusters, but require an enhanced $s$-processes.

The O, Ca, K, and Cu residuals are more surprising, as the prior open cluster studies do not identify enhancements in these elements (or do not observe these elements). Our Ca residuals appear robust, as five of the six youngest clusters show a positive Ca residual of 0.15-0.25 dex that falls outside or on the upper edge of field sample residuals. We see the largest residual abundances for K, with five of the six youngest clusters displaying $\Delta$[K/H] of $\sim$0.25 dex or greater. The median Cu residuals are less uniform. Among the the clusters with $\geq4$ stars with unflagged Cu abundances, four have $\Delta$[Cu/H] near zero (tend to be older) and six have residuals near $-0.2$ dex (tend to be younger). We note that \citet{casamiquela2021} see depletion in [Zn/H] in Gaia ESO clusters, an element with similar nucleosynthetic origin to Cu, though we find [Zn/H] abundance residuals near zero.

For O, all clusters show residuals $>0.05$ dex. This is surprising, but not entirely unexpected given that the field star comparison for clusters with high median [Fe/Mg] and low median [Mg/H] also show O enhancements (e.g., NGC 2232, ASCC 16), indicating potential bias in our O residuals in this abundance space. As discussed in Section~\ref{subsec:As}, stars in this metallicity range may suffer from rotationally broadened lines and have artificially low [Mg/H] values that drive high $\AIa/\Acc$ and poor two-process fits. However, Figure~\ref{fig:cluster} positive $\Delta$[O/H] for all clusters, regardless of median [Mg/H] and [Fe/Mg]. 

The clear trends in abundance residuals with age for O, Ca, K, and Cu as well as the enhancements in Y and Ba show that young open clusters have unique chemical enrichment and that abundance residuals are correlated with age. The enhancement of O, Ca, and K and the depletion of Cu have not previously been identified in cluster surveys and should be studied further. We find a distinct residual abundance pattern for each cluster, which is encouraging for chemical tagging, though it is unclear from our current sample if clusters of the same age could be distinguished with this method.  

\section{Adding an AGB process} \label{sec:agb}

The prior sections of this paper have focused on CCSN and SNIa enrichment, the two main producers of lighter elements. However, heavy elements such as Y and Ba are predominantly produced through slow neutron-capture nucleosynthesis \citep{arlandini1999, bisterzo2014} in neutron rich environments, such as AGB stars \citep[e.g.,][]{karakas2016}, and they are expected to have little or no SNIa contribution. To better describe these two elements, we add a third AGB component to and construct the three-process model:
\begin{equation}\label{eq:xh_3}
    \xh = \log_{10}[\Acc\qccx(z) + \AIa\qIax(z) + \Aagb\qagbx(z)].
\end{equation}
Fitting a general three-process model is challenging because SNIa and AGB enrichment are both delayed in time, and without a detailed theoretical prior on yields there is no obvious way to separate them. In this paper we adopt a ``restricted'' three process model by setting $\qagbx=0$ for elements O to Cu, and $\qIax=0$ for Y and Ba. Some other elements in our data set may have non-zero AGB contributions, and we can examine this to some degree by checking whether their two-process residuals correlate with $\Aagb$. 

\subsection{Fitting the AGB process vectors}\label{subsec:agb_fit}

We preserve the two-process $\qccx$ and $\qIax$ vectors for the light elements and define $\qagbx$ and $\qccx$ for Ba and Y based off of the median [Ba/Mg] and [Mg/H] trends. We choose to model the AGB component with Ba because it is better measured by GALAH than Y. With $z = 10^{\mgh}$, $z_{\rm Ba} = 10^{[\rm Ba/H]}$, and $\AIa^{\rm Ba}=0$, Equation~\ref{eq:xh_3} reduces to
\begin{equation}  \label{eq:z_ba}
    z_{\rm Ba} \equiv 10 ^{\rm [Ba/H]} = \Acc^{\rm Ba}\qcc^{\rm Ba}(z) + \Aagb^{\rm Ba}\qagb^{\rm Ba}(z)
\end{equation}
for Ba, resembling the two-process model for Fe. Inferring the metallicity dependence of $\qagb^{\rm Ba}$ from median trends requires \textit{some} assumption about the metallicity dependence of $\qcc^{\rm Ba}$, and the observed trends (Figure~\ref{fig:xmgs}) suggest that the metallicity dependence for AGB enrichment is stronger. Though not necessarily correct, we assume $\qcc^{\rm Ba}$ to be metallicity independent. As with [Fe/Mg], we assume that the low metallicity, low-Ia plateau at [Ba/Mg]$=-0.4$ represents pure CCSN enrichment with $\Aagb=0$. Thus,

\begin{equation}\label{eq:q_cc_ba}
    \qcc^{\rm Ba} = ({\rm Ba/Mg})_{\rm pl} = \frac{\Acc\qcc^{\rm Ba}}{\Acc\qcc^{\rm Mg}} = 10^{[{\rm Ba/Mg}]_{\rm pl}},
\end{equation}
where we have used the two-process assumption that $\qcc^{\rm Mg}=1$.
At solar abundances we have $\Acc = \Aagb = z = z_{\rm Ba} = 1$ by definition. Since we are assuming $\qIa^{\rm Ba}=0$ we must have $\qcc^{\rm Ba} + \qagb^{\rm Ba}=1$ at solar, implying
\begin{equation}
    \qagb^{\rm Ba}(z=1) = 1 - \qcc^{\rm Ba} = 1 - 10^{[\rm Ba/Mg]_{\rm pl}}.
\end{equation}

At other metallicities, we have $\Acc$ from the two-process fit as well as the median low-Ia and high-Ia $z_{\rm Ba}$ for each $z$ bin. With these assumptions and constraints we are left with three unknowns: $\qagb^{\rm Ba}$, $\Aagb$ on the low-Ia sequence, and $\Aagb$ on the high-Ia sequence. In the two-process model for Mg and Fe, we reduce the number of unknowns to two by assuming that $\qIa^{\rm Fe}$ is metallicity independent. This allows us to to infer $\AIa$ on both sequences. While this is a reasonable assumption for Fe, $\qagb^{\rm Ba}$ should be metallicity dependent \citep[e.g.,][]{cristallo2011, karakas2016}. In our restricted three-process model, we instead assume that $\Aagb = \AIa$ on the high-Ia sequence, where the delayed contribution dominates. At [Mg/H]=0 this is true by definition, so we are simply using the inferred value at $\AIa$ as our best estimate of $\Aagb$ at other metallicities. From these assumptions we can now infer $A_{\rm AGB, \, low}$ and $\qagb^{\rm Ba}(z)$ from $z_{\rm Ba, \, low}$ and $z_{\rm Ba, \, high}$, where low and high refer to the low-Ia and high-Ia populations. 

Re-writing Equation~\ref{eq:z_ba} for both populations, we find
\begin{equation}\label{eq:z_ba_high}
    z_{\rm Ba, \, high} = A_{\rm cc, \, high}\qcc^{\rm Ba} + A_{\rm AGB, \, high} \qagb^{\rm Ba}(z),
\end{equation}
\begin{equation}\label{eq:z_ba_low}
    z_{\rm Ba, \, low} = A_{\rm cc, \, low}\qcc^{\rm Ba} + A_{\rm AGB, \, low} \qagb^{\rm Ba}(z).
\end{equation}
With $\qcc^{\rm Ba}$ as defined in Equation~\ref{eq:q_cc_ba} and $A_{\rm AGB, \, high} = A_{\rm Ia, \, high}$, we can solve Equation~\ref{eq:z_ba_high} for $\qagb^{\rm Ba}(z)$ as
\begin{equation}
    \qagb^{\rm Ba}(z) = \frac{z_{\rm Ba, \, high} - A_{\rm cc, \, high} 10^{\rm [Ba/Mg]_{pl}}}{A_{\rm Ia, \, high}},
\end{equation}
where $A_{\rm cc, \, high}$ and $A_{\rm Ia, \, high}$ are the median process amplitudes on the high-Ia sequence from the Fe and Mg two-process fit.
This allows us to solve Equation~\ref{eq:z_ba_low} for $A_{\rm AGB, \, low}$, such that
\begin{equation}
    A_{\rm AGB, \, low} = \frac{z_{\rm Ba, \, low} - A_{\rm cc, \, low} 10^{\rm [Ba/Mg]_{pl}}} {\qagb^{\rm Ba}(z)}.
\end{equation}

We now know the process amplitudes along the high-Ia and low-Ia sequences. Generalizing Equations~\ref{eq:z_ba_high} and~\ref{eq:z_ba_low} to any other element with $\qIax=0$, we then solve the system of equations for $\qccx$ and $\qagbx$:
\begin{equation}\label{eq:q_cc_agb}
    \qccx(z) = \frac{z_{X \rm , \, high} A_{\rm AGB, \, low} - z_{X \rm , \, low} A_{\rm AGB, \, high}}{A_{\rm cc, \, high}A_{\rm AGB, \, low} - A_{\rm cc, \, low}A_{\rm AGB, \, high}}
\end{equation}
\begin{equation}\label{eq:q_agb}
    \qagbx(z) = \frac{z_{X \rm,\,low} - A_{\rm cc, \, low}\qccx}{A_{\rm AGB, \, low}}.
\end{equation}
In this paper, the only other element we fit in this way is Y. 

\begin{figure}[!htb]
    \centering
    \includegraphics[width=\columnwidth]{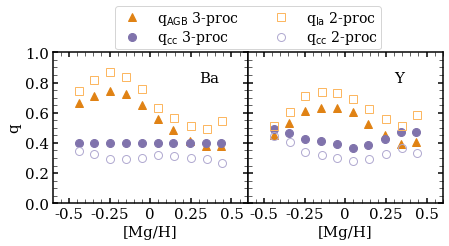}
    \caption{Process vectors $\qagb$ (dark orange triangles) and $\qcc$ (dark purple circles) for the restricted three-process fit to Ba (left) and Y (right). The two-process $\qIa$ (light orange open squares) and $\qcc$ (light purple open circles) values are plotted for comparison.}
    \label{fig:qs_agb}
\end{figure}

In summary, we infer the AGB process vectors $\qagb^{\rm Ba}(z)$ and $\qagb^{Y}(z)$ from the observed median [Ba/Mg] and [Y/Mg] sequences of the low-Ia and high-Ia populations, using Ba as the tracer of AGB enrichment analogusly to the way we use Fe in determining $\qIax(z)$. The fitting requires some additional assumptions, so the inferred metallicity dependence should be taken as approximate. We plot the three-process $\qcc$ and $\qagb$ vectors for Ba and Y in Figure~\ref{fig:qs_agb} alongside $\qcc$ and $\qIa$ from the two-process model (Section~\ref{subsec:qs}). 

With the process vectors defined, we can infer the $\Aagb$ of individual stars. In the two-process model we use a weighted mixture of six elements and $\chi^2$ minimization to find the best process amplitudes for each star. Here we only have two reliable neutron-capture elements, and Ba measurements are usually more precise than Y measurements. We therefore take a simpler approach and estimate $\Aagb$ from Ba alone,
\begin{equation}
    \Aagb^{\rm Ba} = \frac{z_{\rm Ba} - \Acc^{\rm Ba}\qcc^{\rm Ba}}{\qagb^{\rm Ba}},
\end{equation}
where we take $\Acc$ from the two-process model and interpolate $\qccx$ and $\qagbx$ to the star's value of $\mgh$. We have also tried calculating $\Aagb$ from a weighted sum of Ba and Y and found similar results. 

\subsection{Application}\label{subsec:agb_ap}

\begin{figure*}[!htb]
    \centering
    \includegraphics[width=\textwidth]{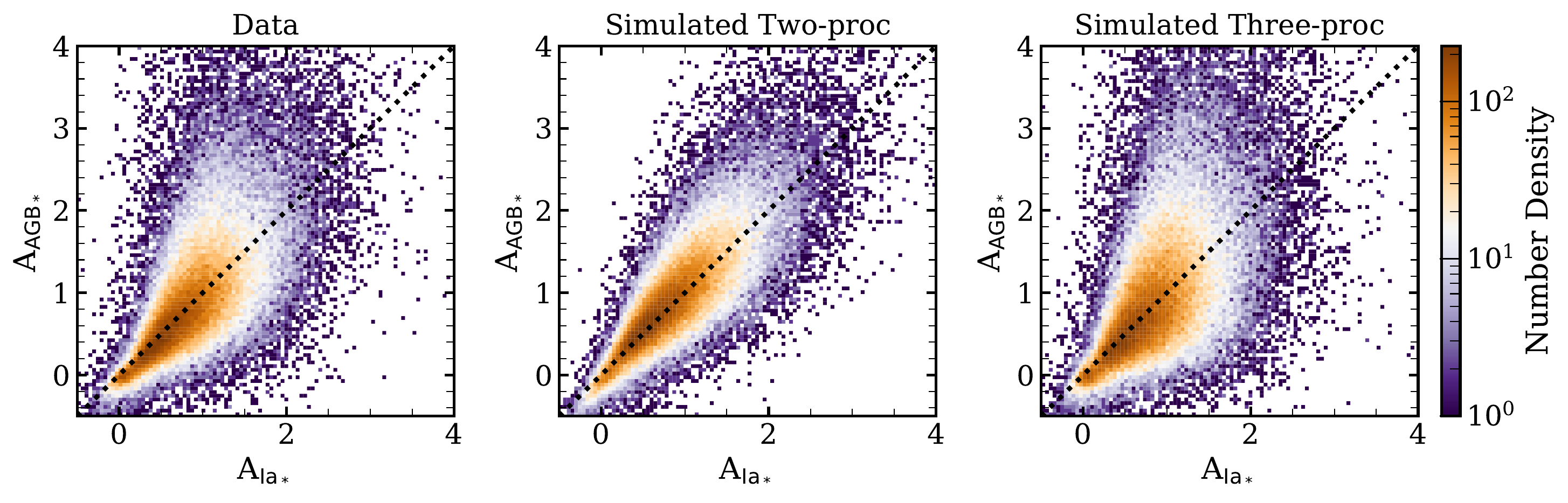}
    \caption{A comparison of stellar $\Aagb$ and $\AIa$ values in a 2D density histogram, with orange tones representing a high number density and purple tones a low number density. The dotted black line indicates a 1:1 relationship. The distribution is shown for the amplitudes fit to the GALAH abundances (Left), the simulated two-process abundances (Center), and the simulated three-process abundances (Right).}
    \label{fig:Aagb}
\end{figure*}

\begin{figure*}[!htb]
    \centering
    \includegraphics[width=\textwidth]{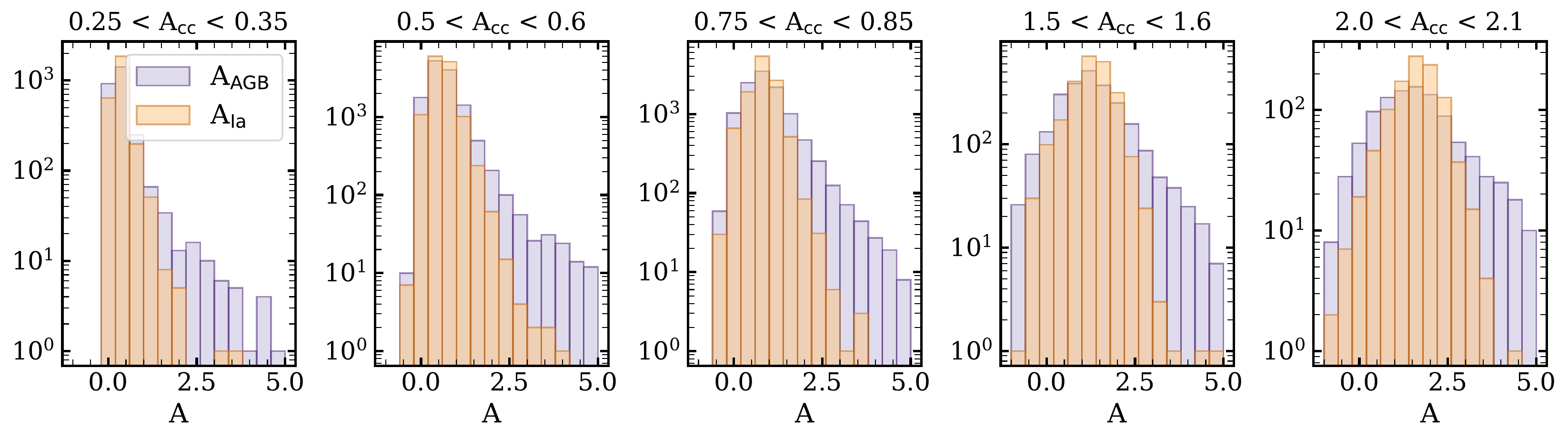}
    \caption{Distribution of $\AIa$ and $\Aagb$ values for five 0.1 ranges of $\Acc$ (starting at 0.25, 0.5, 0.75, 1.5, and 2.0, increasing to the right). $\Aagb$ distributions are colored purple and $\AIa$ distributions are colored orange. Note that the y-axes range varies between panels.}
    \label{fig:A_dist}
\end{figure*}

Our first goal is to determine if the AGB process is observably distinct from the SNIa process. In the left panel of Figure~\ref{fig:Aagb}, we plot $\AIa$ from the two-process fit vs. $\Aagb$ from the three-process fit. We see that the core of the distribution follows a one-to-one relationship, indicating that the two-process model and the three-process model find similar amplitudes of a delayed process for most stars. However, we also find a population of stars best fit with $\AIa$ between 1 and 2.5 in the two-process model that are fit by a larger $\Aagb$ in the three-process model. To understand how $\Aagb$ and $\AIa$ vary with $\Acc$, we include distributions of the stellar amplitudes in 0.1 bins of $\Acc$ in Figure~\ref{fig:A_dist}. We show five ranges of $\Acc$, from 0.25 to 2.1. As $\Acc$ increases, the peak in the $\AIa$ and $\Aagb$ distributions also increases. As in Figure~\ref{fig:Aagb}, the $\Aagb$ values span a larger range than $\AIa$ in every $\Acc$ bin. While few stars exceed $\AIa$ of 2.5, a large number of stars have an $\Aagb$ value of 2.5-5.

The stars that are fit with large $\Aagb$ have high [Ba/H] ($\sim 0.3-0.6$). To test if this population could emerge from observational scatter, we create a simulated three-process data set similar to the simulated two-process data set described in Section~\ref{sec:pred&devs} but with Ba and Y abundances based on the three-process fit to each sample star. As previously, we add random errors to the Ba and Y abundances drawn from a Gaussian distribution with the reported GALAH uncertainty. The middle and right panels of Figure~\ref{fig:Aagb} show the $\Aagb-\AIa$ distribution of the simulated two-process and three-process data sets, respectively. The simulated three-process distribution agrees well with the observed distribution, implying that the broad range of $\Aagb$ at fixed $\AIa$ is consistent with that expected from observational errors. Conversely, the simulated two-process data set does not reproduce the observed upturn towards high $\Aagb$ at high $\AIa$. This comparison shows that a significant population of sample stars have Ba enrichment substantially above the two-process model prediction, leading to high values of $\Aagb$.

\begin{figure}[!htb]
    \centering
    \includegraphics[width=\columnwidth]{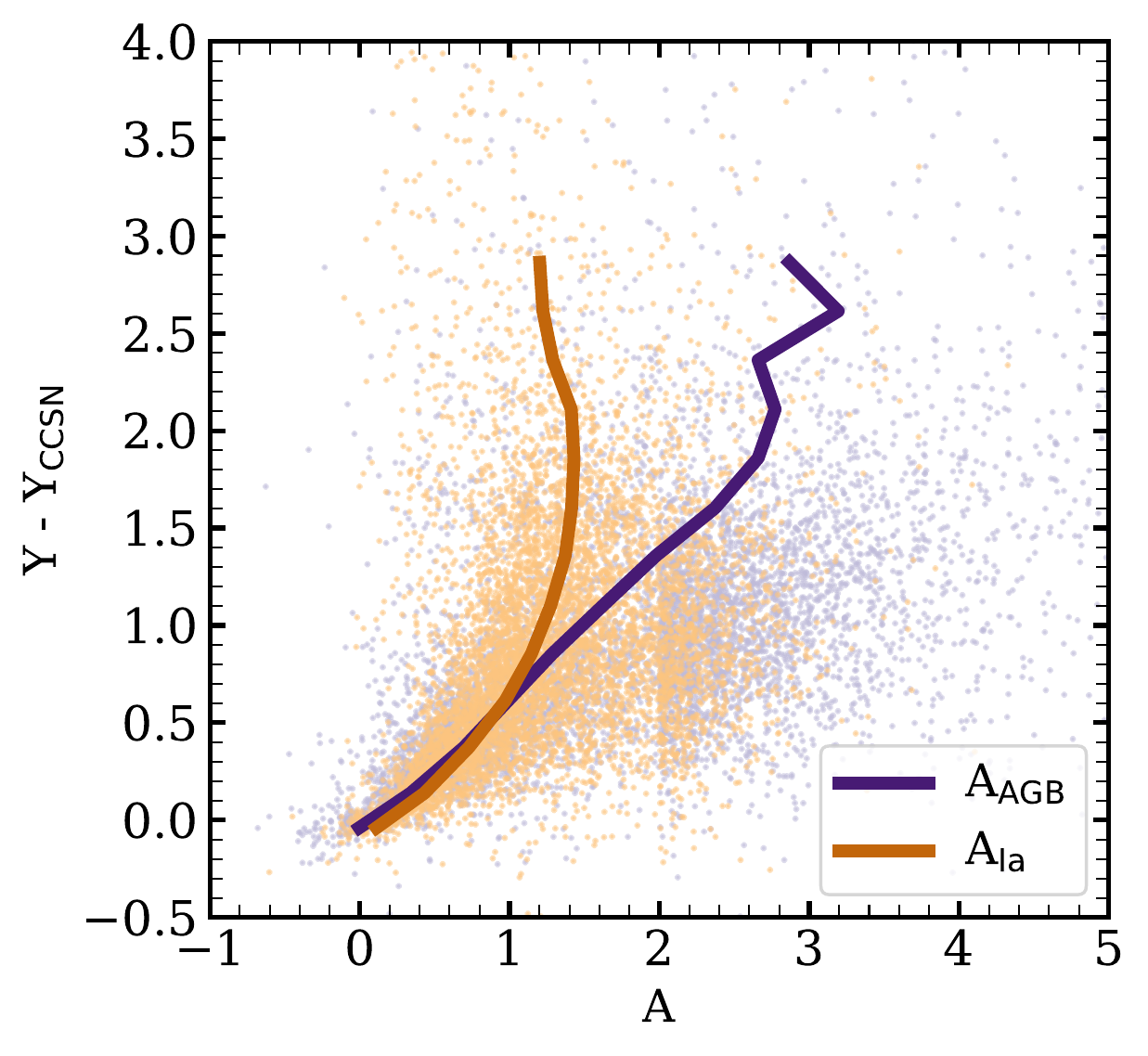}
    \caption{Process amplitude vs. non-CCSN Y (Equation~\ref{eq:Yagb}) in solar abundances for $\AIa$ (orange) and $\Aagb$ (purple). Star are down-sampled to 4\% of the population at $A<2$ and $\rm{Y} - \rm{Y_{\rm cc}}<1.5$ and 40\% elsewhere. The median $\AIa$ and $\Aagb$ in bins of $\rm{Y} - \rm{Y_{\rm cc}}$ are plotted as solid orange and purple lines, respectively. }
    \label{fig:Y_agb}
\end{figure}

If the AGB amplitude fit to Ba indeed represents AGB enrichment, then Y abundances should also track $\Aagb$ more closely than they track $\AIa$. For this test, we wish to remove the CCSN contribution to Y. Figure~\ref{fig:Y_agb} plots the stellar $\AIa$ (orange) and $\Aagb$ (purple) amplitudes against the non-CCSN component of Y in solar abundances:
\begin{equation}\label{eq:Yagb}
    {\rm Y} - {\rm Y_{\rm cc}} = 10^{\rm [Y/H]} - \Acc \qcc^{\rm Y}.
\end{equation}
Stars are down-sampled such that we show 4\% of the population at $A<2$ and $\rm{Y} - \rm{Y_{\rm cc}}<1.5$ and 40\% elsewhere. We plot the median $A$ in bins of $\rm{Y} - \rm{Y_{\rm cc}}$ for both the SNIa and AGB amplitudes. Below $A=1$, the non-CCSN component of Y increases with increasing amplitude, and the AGB and SNIa amplitudes track each other. For $A>1$ we find a large amount of star-to-star scatter, but distinct median trends. While the median $\AIa$ saturates around 1.4 for high Y values, the median $\Aagb$ continues to increase for rising $\rm{Y} - \rm{Y_{\rm cc}}$. This trend shows that large $\Aagb$ values do successfully predict a large non-CCSN contribution to Y, even though the observational errors in Y and Ba abundances are large enough to add substantial scatter to this relation. While the delayed SNIa and AGB processes approximately track each other, leading to large separations in the [Ba/Mg] and [Y/Mg] trends of low-Ia and high-Ia stars (Figure~\ref{fig:xmgs}), Figures~\ref{fig:Aagb} to~\ref{fig:Y_agb} support the expectation that Ba and Y enrichment arises from a physically distinct process, and they show that some stars have AGB enrichment well above that of typical stars with the same $\Acc$ and $\AIa$. 

As done in Section~\ref{subsec:age} for two-process model residuals, we investigate the trend of AGB enrichment with age for high-Ia stars. To further understand the differences between the two and three-process models' SNIa and AGB enrichment with age, we construct a new parameter $\Delta\Aagb$,
\begin{equation}\label{eq:delta_agb}
    \Delta\Aagb = \Aagb - {\rm median}(A_{\rm AGB, \, binned}),
\end{equation}
where ${\rm median}(A_{\rm AGB, \, binned})$ is the median $\Aagb$ value in bins of $\AIa/\Acc$ with a width of 0.3. $\Delta\Aagb$ describes a star's AGB enrichment relative to other stars with similar SNIa enrichment. If a star has more AGB enrichment than predicted by its $\AIa/\Acc$, $\Delta\Aagb$ will be positive. A $\Delta\Aagb$ of 1, for example, would imply that the AGB enrichment is twice solar for an otherwise solar star.

\begin{figure}[!htb]
    \centering
    \includegraphics[width=\columnwidth]{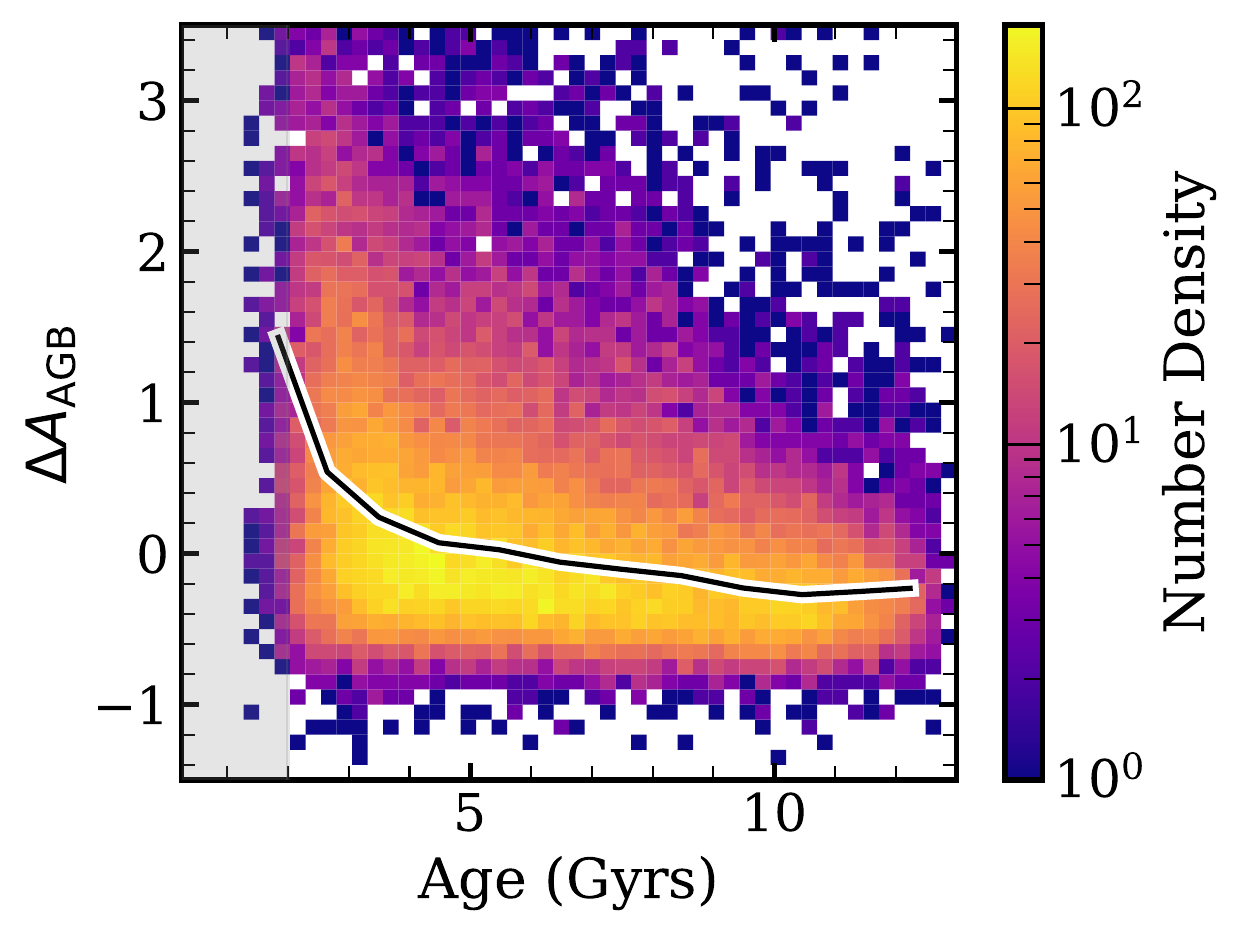}
    \caption{2D histogram showing the correlation of $\Delta\Aagb$ (Equation~\ref{eq:delta_agb}) and stellar age for high-Ia stars with [Mg/H]$>-0.3$. Yellow tones represent high number density. We plot the median $\Delta\Aagb$ value in 1 Gyr bins of age as a solid black line. We shade the region from 0 to 2 Gyrs grey to indicate uncertain ages, as in Figure~\ref{fig:delta_age}}
    \label{fig:agb_age}
\end{figure}

We plot the distribution of $\Delta\Aagb$ with age for high-Ia stars with [Mg/H]$>-0.3$ in Figure~\ref{fig:agb_age} along with the median $\Delta\Aagb$ in bins of age. We see that a significant fraction of stars have high $\Delta\Aagb$ ($>0.5)$. In an average sense, the youngest stars tend to have more AGB enrichment relative to SNIa, though the scatter in the distribution grows as the age decreases. This relationship is a statement about Ba enrichment, since we derive $\Aagb$ from Ba alone, and it agrees with the Ba age trends seen in Figure~\ref{fig:delta_age}. Y follows the same trend, but it is not as well measured by GALAH. The conclusion that AGB enrichment to Ba and Y is time dependent and distinct from SNIa agrees with observations of neutron-capture enrichment in open clusters (see Section~\ref{sec:clusters}) and Ce trends in \citetalias{weinberg2021}.

Several of the other elements that exhibit large separations between low-Ia and high-Ia median trends could plausibly have AGB contributions, such as Na, Cu, and Zn. We have looked for correlations between two-process residuals for these elements and values of $\Aagb/\AIa$, but we find no clear signal. This lack of trend could be an indication that the AGB contribution to these elements is small, or it could simply indicate that the observational errors in GALAH DR3 remain too large to disentangle SNIa and AGB contributions. \citetalias{weinberg2021} discuss the general problem of describing abundance patterns with an $N$-process model (see their Section 8), but because their APOGEE data set has only one neutron-capture element (Ce) they do not attempt a three-process decomposition like the one undertaken here. We hope that future data sets with more high quality neutron-capture abundances, plus guidance from simulations that incorporate multiple enrichment channels, will lead to further progress in constraining the contributions of AGB and other processes to multi-element abundance distributions.

\section{Summary} \label{sec:conclusions}

We present the [X/Mg] vs. [Mg/H] abundance trends and two-process residual abundances for 82,910 stars in GALAH DR3. Our stars span an abundance range of $-0.5 < \mgh < 0.5$ and are restricted to dwarf and subgiant stars ($3.5< \logg < 4.5$) with temperatures $4200 \kel < \teff < 6700 \kel$ to avoid systematic abundance trends that correlate with stellar parameters. Our abundance sample includes $\alpha$ (Mg, O, Si, Ca, Ti), light odd-$Z$ (Na, Al, K, Sc), Fe-peak (Cr, Mn, Fe, Ni), Fe-cliff (Cu, Zn) and neutron-capture (Y, Ba) elements that span CCSN, SNIa, and AGB production \citep{andrews2017, rybizki2017}. 

\textbf{Median Trends}: We divide the sample into high-Ia and low-Ia populations, as in \citet{weinberg2019}, \citet{griffith2019}, and \citetalias{weinberg2021}, and compare the median high-Ia and low-Ia trends for GALAH DR3 with those from GALAH DR2 and APOGEE DR17. We observe the following: 
\begin{enumerate}
\itemsep0em
    \item The GALAH DR2 and D3 median trends are consistent for most elements. The K, Y, and Ba trends change the most, likely due to new NLTE corrections for K and differences in the abundance analysis between DR2 and DR3.
    \item There is little to no separation in the median trends of O, Si, Ca, Ti, Al, K, Sc, and Zn, suggesting that prompt CCSN dominate the production of these elements. Conversely, Na, Cr, Mn, Fe, Ni, Cu, Y, and Ba have significant separation between their median high-Ia and low-Ia trends and thus have large delayed contribution from SNIa or AGB stars.
    \item As in DR2, the metallicity dependent [O/Mg] trends in GALAH DR3 are in strong disagreement with the flat APOGEE trends, though both surveys find no separation in the O median sequences. 
    \item This work adds a comparison of GALAH Cu and Ba trends to APOGEE DR16 Cu and DR17 Ce, respectively. There is good agreement in the surveys' Cu trends above $\mgh=0$ and of the Ba and Ce trends, which display the same peaked metallicity dependence in their high-Ia sequences.
    \item The separation between the high-Ia and low-Ia medians is larger for the GALAH abundance trends than those of APOGEE at $\mgh>0.25$ for Si, Ca, Cr, Mn, Fe, and Ni. This could be explained by differences in the APOGEE and GALAH samples or observational uncertainties. 
\end{enumerate}

\textbf{Two-process model}: We then fit our sample with the two-process model, deriving process vectors ($\qcc$ and $\qIa$) from the median high-Ia and low-Ia sequences and fitting process amplitudes ($\Acc$ and $\AIa$) to each star from its Mg, Si, Ca, Ti, Fe, and Ni abundances. From the process vectors we derive the fractional CCSN contribution to stars on the high-Ia sequence at solar metallicity (see Table~\ref{tab:elems}), finding values in agreement with the qualitative description in point 2 above. With the process vectors and amplitudes known, we predict the full suite of elemental abundances from Equation~\ref{eq:xh} for each star in our sample. We then analyze the residual abundances -- the differences between the observed and predicted values. As in \citetalias{weinberg2021}, we find that the two-process model is a better predictor of observed abundances than the median trends. 

\textbf{Residual Abundances}: We compare the two-process residual abundances with residuals for a simulated abundance set, which we construct by assigning stars the two-process abundance value plus a random error from a Gaussian distribution resembling the reported GALAH abundance errors. Stars in the core of the residual abundance distributions are well predicted by the two-process model and can be explained by Gaussian observational noise. Stars in the wings of the residual abundance distributions cannot be explained by Gaussian noise alone, and instead may have true deviations from the two-process model or larger observational errors. To identify underlying structure in the residual abundances, we compute the covariance of the intrinsic (data - simulation) residual abundances. In almost all cases the simulated covariance is smaller than that of observed data's covariance. We find that the Ba, Y, and Zn residuals are all positively correlated, suggestive of AGB enrichment. For the lighter elements, the correlated patterns are less clear than those found by \citetalias{weinberg2021} in APOGEE data, perhaps because of larger observational errors to residual abundances.

We identify correlations in the residual abundances with age \citep{sharma2020} for O, K, Y, and Ba. The wings of the K residual abundance distribution increase to $\sim0.8$ dex at young ages, but the core of the distribution remains uncorrelated with age. Conversely, the core of the O, Y, and Ba residual abundance-age distribution rises with decreasing age for stars $\lesssim3$ Gyrs. The Y and Ba age trends are consistent with those found for Ce in \citetalias{weinberg2021} and with enhancements of neutron-capture elements in young open clusters, but the O and K trends are more difficult to interpret. Because our sample includes rapidly rotating young stars, rotational broadening and resulting poor abundance determination may skew the observed trends. 

\textbf{Large Deviations}: To determine if the stars on the extreme tails of the residual abundance distribution have real enhancements/depletions relative to the two-process model, we inspect the spectra of 100 stars in the 99th percentile of the $\chi^2$ distribution. From this analysis we conclude that roughly 40$\%$ of highly deviating stars have real residual abundances, while the remaining 60$\%$ suffer from problematic data and have no genuine deviations. We identify 22 stars with broad or double peaked lines that may be unflagged binaries. 

Seven of the 100 inspected stars have interesting and robust residual abundances with no indication of observational systematics. We compute the SME line profile predicted by the two-process abundances for these stars and confirm that they do not match the observed spectral features. One such star displays a large Na residual ($\sim0.5$ dex), but is otherwise well predicted by the two-process model. In the full stellar sample, we identify a total of 15 salty stars with $\Delta$[Na/Mg]$>0.3$ dex and small ($<0.15$ dex) deviations for elements O through Ni. Interestingly, many of these stars have positive Cu, Zn, Y, and Ba residuals as well. While data systematics afflict many of the stars in the 99th percentile of the $\chi^2$ distribution, those with real residual abundances can lead us to populations of peculiar abundance stars worthy of further study. 

\textbf{Open Clusters:} We analyze the residual abundances of 14 open clusters, taking cluster members and ages from \citet{spina2021}. Because open clusters should be chemically homogeneous (at the level of GALAH abundance precision), we take the median abundance of cluster residuals to reduce the impact of statistical uncertainties and systematic errors. We find that young open clusters are enhanced in O, Ca, K, Y, and Ba and depleted in Cu, with the magnitude of residuals strongly correlated with age. The enhancement of neutron-capture elements Y and Ba in young open clusters is in agreement with past works \citep[e.g.,][]{dorazi2009}, but the O, Ca, K, and Cu trends are more surprising. The residual abundance-age trends in GALAH open clusters are a promising sign for chemical tagging and reveal the two-process model's power in identifying the interesting abundance trends hidden beneath the global enrichment patterns.

\textbf{Three-process Model:} We leverage the lack of SNIa enrichment to Y and Ba to construct a restricted three-process model for CCSN, SNIa, and AGB enrichment. This model is restricted as it assumes $\qIa=0$ for Y and Ba and $\qagb=0$ for all other elements. We derive CCSN and AGB process vectors and amplitudes from the Mg and Ba trends. Through a comparison of $\AIa$ and $\Aagb$, we determine that the AGB process is distinct from the SNIa process and that the AGB process, fit to Ba, can better reproduce the Y enrichment than the SNIa process, fit to Fe. We also identify a population of stars with AGB enrichment substantially above the average level predicted by their SNIa enrichment. This population is more prevalent at ages $\leq 3$ Gyrs. From our analysis we conclude that Y and Ba are enriched by a distinct, non-SNIa source. Although this conclusion is theoretically unsurprising, it is challenging to demonstrate from abundance trends alone.

\textbf{Future work:} This work complements \citetalias{weinberg2021} and serves as further proof of concept that the two-process residual abundances hold a wealth of information that can identify unique stars/populations of stars and non-CCSN/SNIa enrichment sources. Our analysis in Sections~\ref{sec:pred&devs} to~\ref{sec:agb} provide initial illustrations of what can be done with these residuals. Future works should follow up interesting trends identified in this paper (e.g., Na-rich stars and open cluster O, Ca, K, Cu enrichment) and conduct a more complete search for robust residual abundances among stars that deviate strongly from the two-process predictions. We plan to continue studying Galactic evolution with the two-process model and residual abundances. As spectroscopic surveys such as Milky Way Mapper \citep{kollmeier2017} expand the coverage of our Galaxy and nearest neighbors, and as abundance pipelines achieve higher levels of accuracy and precision, we will improve our understanding of our Galactic chemical enrichment and the astrophysical origins of the elements.

\section*{Acknowledgement}

This work is supported by NSF grant AST-1909841.
DHW gratefully acknowledges the hospitality of the IAS and financial support of the W.M. Keck and Hendricks Foundations during much of this work. 

This work made use of the Third Data Release of the GALAH Survey (Buder et al. 2021). The GALAH Survey is based on data acquired through the Australian Astronomical Observatory, under programs: A/2013B/13 (The GALAH pilot survey); A/2014A/25, A/2015A/19, A2017A/18 (The GALAH survey phase 1); A2018A/18 (Open clusters with HERMES); A2019A/1 (Hierarchical star formation in Ori OB1); A2019A/15 (The GALAH survey phase 2); A/2015B/19, A/2016A/22, A/2016B/10, A/2017B/16, A/2018B/15 (The HERMES-TESS program); and A/2015A/3, A/2015B/1, A/2015B/19, A/2016A/22, A/2016B/12, A/2017A/14 (The HERMES K2-follow-up program). We acknowledge the traditional owners of the land on which the AAT stands, the Gamilaraay people, and pay our respects to elders past and present. This paper includes data that has been provided by AAO Data Central (datacentral.aao.gov.au).

This work has made use of data from the European Space Agency (ESA) mission Gaia (https://www.cosmos.esa.int/gaia), processed by the Gaia Data Processing and Analysis Consortium (DPAC, https://www.cosmos.esa.int/web/gaia/dpac/consortium). Funding for the DPAC has been provided by national institutions, in particular the institutions participating in the Gaia Multilateral Agreement.

\appendix
\section{Carbon} \label{ap:carbon}

Though not discussed in the main text, we are interested in the C abundance trends because of its debated nucleosynthetic origin. C production through the triple-$\alpha$ process \citep{salpeter1952} has known production in massive stars with strong ties to wind and rotation \citep[e.g.,][]{gustafsson1999, meynet2002}. Recent observations \citep[e.g.,][]{bensby2006,cescutti2009,nissen2014} find evidence of substantial additional C production in low-intermediate mass AGB stars. The relative contribution of massive stars and AGB stars to C production remains uncertain. 

\begin{figure*}[!htb]
    \centering
    \includegraphics[width=\textwidth]{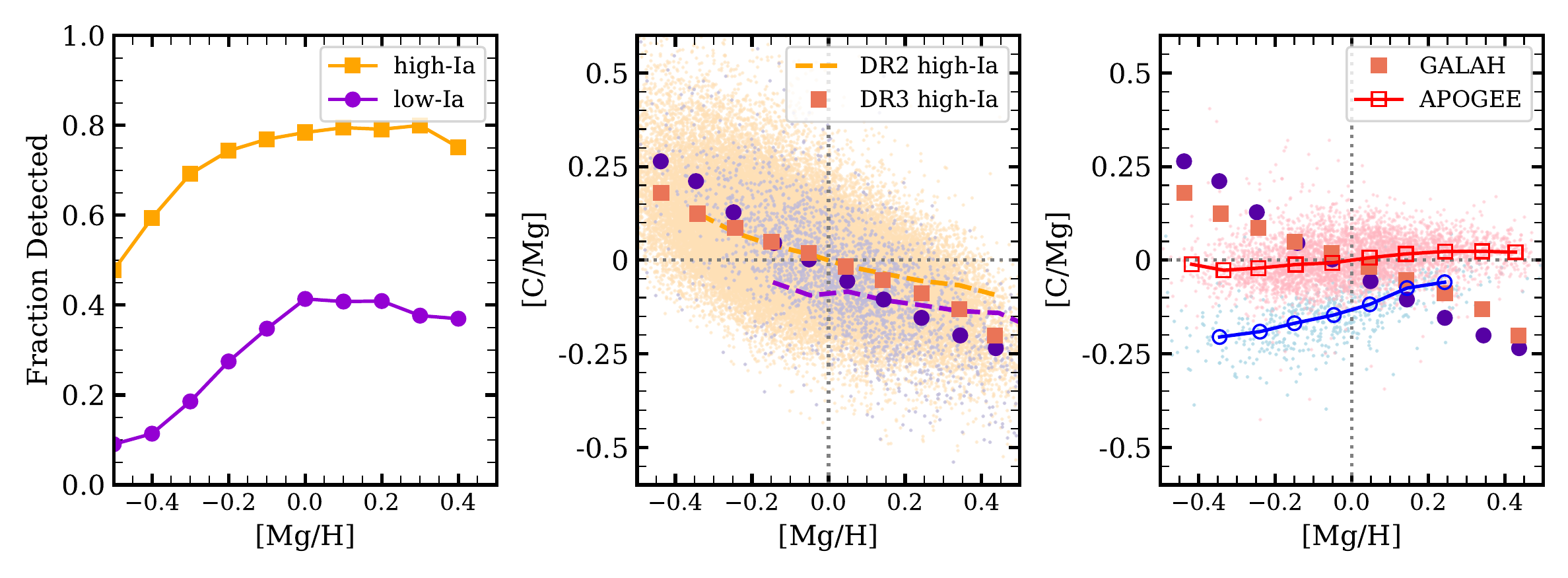}
    \caption{\textit{Left:} Fraction of stars with a C detection in our high-Ia (magenta) and low-Ia (orange) populations. \textit{Middle:} Same as Figure~\ref{fig:xmgs}, but for C. \textit{ Right:} Same as Figure~\ref{fig:xmgs_apogee}, but for C. Here, APOGEE C abundances are those corrected to the birth abundance from \citep{vincenzo2021c}}
    \label{fig:cmg}
\end{figure*}

We plot the median [C/Mg] vs. [Mg/H] trends for GALAH DR3 (solid points) abundances in the middle panel of Figure~\ref{fig:cmg} (with an offset of -0.2 dex). Though C is produced by a mix of prompt (CCSN) and delayed (SNIa) sources, we find that the median high-Ia and low-Ia trends overlap. At face value this would suggest that prompt enrichment dominates C production, however, the median trends are likely skewed by the detectability of C (left panel of Figure~\ref{fig:cmg}). As discussed in Section~\ref{sec:data}, we exclude C from the main analysis because it is difficult to determine in GALAH, which observes one atomic C line. GALAH reports a C abundance for less than 80\% of the high-Ia stars and less than 40\% of the low-Ia stars, with detection fractions dropping as low as 10\% at low metallicity. The low detection fraction implies that a large number of stars with [C/Mg] below the detection threshold are not included in the median calculation. If these low abundance stars (or their upper limits) were included, it would likely drive the median trends down. We expect that the low-Ia median would drop significantly, and that the metallicity dependence would flatten at [Mg/H]$<0$ where the detection fraction drops. 

The median high-Ia and low-Ia [C/Mg] vs. [Mg/H] trends are closer together and more inclined in GALAH DR3 than in DR2 \citep[middle panel Figure~\ref{fig:cmg},][]{griffith2019}. This change results from the switch away from data-driven models that imposed abundance trends from the training set onto the global population. We are skeptical of interpreting the DR2 or DR3 trends given the low detection fraction and difficulty in determining robust C abundances. In the final panel of Figure~\ref{fig:cmg} we plot the median C abundance trends from GALAH DR3 and APOGEE DR16. The APOGEE C abundances are taken from \citet{vincenzo2021c}, who apply mixing corrections to infer birth abundances of C and N for a sample of red giants with asteroseismic masses. We apply zero point offsets to both data sets to ensure that the high-Ia sequence passes through [C/Mg]=0 at [Mg/H]=0. The [C/Mg] vs. [Mg/H] trends clearly disagree in both metallicity dependence and sequence separation. The low C detection fraction in GALAH likely inflates the median values at low metallicity, driving the subsolar differences with APOGEE, but it cannot account for the discrepancy at [Mg/H]$>0$. The stark difference in metallicity dependence is reminiscent of the [O/Mg] vs. [Mg/H] trends (Figure~\ref{fig:xmgs_apogee}) for which the APOGEE trends are flat and the GALAH trends are inclined. Further investigation into the GALAH C abundances, such as a determination of upper limits, is required. 

\section{Process Vector Tables}

We include tables of the process vectors $\qcc$ and $\qIa$ for $\alpha$ and light odd-$Z$ elements in Table~\ref{tab:qs1} and Fe-peak, Fe-cliff and neutron-capture elements in Table~\ref{tab:qs2}.

\begin{table}[!htb]
\centering
\caption{The $\qcc$ (top) and $\qIa$ (bottom) values for $\alpha$ and light odd-Z elements in 0.1 dex bins of [Mg/H]. The median [Mg/H] value in each bin is given in the left hand column.} \label{tab:qs1}
\begin{tabular}{rrrrrrrrr}
\hline
\hline
 [Mg/H] &      O &     Si &     Ca &     Ti &     Na &     Al &      K &     Sc \\
\hline
 -0.439 &  1.793 &  0.883 &  0.710 &  0.991 &  0.588 &  0.922 &  0.812 &  0.735 \\
 -0.343 &  1.620 &  0.848 &  0.694 &  0.952 &  0.595 &  0.918 &  0.823 &  0.730 \\
 -0.249 &  1.502 &  0.822 &  0.672 &  0.919 &  0.581 &  0.910 &  0.783 &  0.726 \\
 -0.146 &  1.336 &  0.785 &  0.662 &  0.872 &  0.571 &  0.911 &  0.762 &  0.697 \\
 -0.050 &  1.195 &  0.754 &  0.624 &  0.819 &  0.549 &  0.914 &  0.705 &  0.663 \\
  0.049 &  1.069 &  0.706 &  0.580 &  0.750 &  0.511 &  0.913 &  0.625 &  0.607 \\
  0.145 &  0.932 &  0.643 &  0.550 &  0.661 &  0.426 &  0.867 &  0.530 &  0.529 \\
  0.248 &  0.785 &  0.589 &  0.521 &  0.602 &  0.366 &  0.816 &  0.489 &  0.498 \\
  0.348 &  0.665 &  0.562 &  0.504 &  0.529 &  0.312 &  0.768 &  0.464 &  0.477 \\
  0.439 &  0.549 &  0.529 &  0.486 &  0.487 &  0.426 &  0.760 &  0.463 &  0.457 \\
\hline
 -0.439 & -0.206 &  0.261 &  0.362 &  0.260 &  0.551 &  0.096 &  0.342 &  0.403 \\
 -0.343 & -0.165 &  0.252 &  0.362 &  0.237 &  0.492 &  0.050 &  0.304 &  0.367 \\
 -0.249 & -0.177 &  0.245 &  0.369 &  0.209 &  0.454 &  0.042 &  0.310 &  0.332 \\
 -0.146 & -0.144 &  0.251 &  0.365 &  0.197 &  0.428 &  0.042 &  0.295 &  0.325 \\
 -0.050 & -0.126 &  0.259 &  0.388 &  0.204 &  0.441 &  0.066 &  0.317 &  0.338 \\
  0.049 & -0.126 &  0.285 &  0.413 &  0.232 &  0.498 &  0.106 &  0.363 &  0.393 \\
  0.145 & -0.109 &  0.338 &  0.425 &  0.296 &  0.630 &  0.211 &  0.442 &  0.494 \\
  0.248 & -0.050 &  0.395 &  0.446 &  0.329 &  0.789 &  0.339 &  0.468 &  0.535 \\
  0.348 & -0.023 &  0.416 &  0.464 &  0.395 &  0.978 &  0.476 &  0.482 &  0.552 \\
  0.439 &  0.021 &  0.424 &  0.484 &  0.439 &  0.982 &  0.569 &  0.484 &  0.598 \\
\hline
\end{tabular}
\end{table}

\begin{table}[!htb]
\centering
\caption{Same as Table~\ref{tab:qs1}, but for Fe-peak, Fe-cliff, and neutron capture elements.} \label{tab:qs2}
\begin{tabular}{rrrrrrrrr}
\hline
[Mg/H] &     Cr &     Mn &     Fe &     Ni &     Cu &     Zn &      Y &     Ba \\
\hline
 -0.439 &  0.457 &  0.314 &  0.501 &  0.525 &  0.424 &  0.765 &  0.444 &  0.345 \\
 -0.343 &  0.454 &  0.331 &  0.501 &  0.539 &  0.467 &  0.783 &  0.406 &  0.324 \\
 -0.249 &  0.455 &  0.349 &  0.501 &  0.526 &  0.504 &  0.792 &  0.343 &  0.296 \\
 -0.146 &  0.476 &  0.362 &  0.501 &  0.542 &  0.570 &  0.786 &  0.320 &  0.296 \\
 -0.050 &  0.485 &  0.376 &  0.501 &  0.542 &  0.603 &  0.768 &  0.302 &  0.301 \\
  0.049 &  0.500 &  0.379 &  0.501 &  0.530 &  0.600 &  0.702 &  0.281 &  0.321 \\
  0.145 &  0.530 &  0.392 &  0.501 &  0.496 &  0.546 &  0.595 &  0.293 &  0.312 \\
  0.248 &  0.526 &  0.405 &  0.501 &  0.471 &  0.512 &  0.518 &  0.325 &  0.303 \\
  0.348 &  0.527 &  0.430 &  0.501 &  0.479 &  0.476 &  0.400 &  0.365 &  0.295 \\
  0.439 &  0.572 &  0.507 &  0.501 &  0.499 &  0.499 &  0.415 &  0.333 &  0.266 \\
\hline
 -0.439 &  0.428 &  0.504 &  0.499 &  0.448 &  0.453 &  0.158 &  0.510 &  0.742 \\
 -0.343 &  0.445 &  0.505 &  0.499 &  0.401 &  0.410 &  0.140 &  0.606 &  0.817 \\
 -0.249 &  0.471 &  0.512 &  0.499 &  0.400 &  0.387 &  0.137 &  0.713 &  0.872 \\
 -0.146 &  0.474 &  0.542 &  0.499 &  0.392 &  0.344 &  0.165 &  0.734 &  0.837 \\
 -0.050 &  0.495 &  0.589 &  0.499 &  0.427 &  0.357 &  0.211 &  0.733 &  0.755 \\
  0.049 &  0.521 &  0.655 &  0.499 &  0.500 &  0.438 &  0.322 &  0.690 &  0.633 \\
  0.145 &  0.562 &  0.718 &  0.499 &  0.614 &  0.588 &  0.473 &  0.622 &  0.567 \\
  0.248 &  0.650 &  0.777 &  0.499 &  0.753 &  0.783 &  0.595 &  0.561 &  0.514 \\
  0.348 &  0.751 &  0.826 &  0.499 &  0.870 &  1.030 &  0.745 &  0.513 &  0.492 \\
  0.439 &  0.792 &  0.820 &  0.499 &  0.965 &  1.202 &  0.698 &  0.588 &  0.544 \\
\hline
\end{tabular}
\end{table}

\typeout{}
\bibliography{bibliography}{}
\bibliographystyle{aasjournal}

\end{document}